\documentclass[widetext,showpacs,preprintnumbers,superscriptaddress]{revtex4}

\usepackage[dvipdfm]{graphicx}
\usepackage{amssymb}
\usepackage{amsmath}
\usepackage{subfigure}
\usepackage{dcolumn}
\usepackage{bm}
\usepackage{natbib}
\usepackage{multirow}
\usepackage{aas_macros}

\usepackage{color}

\begin{document}

\preprint{ICRR-Report-608-2011-25,\ IPMU12-0025,\ YITP-12-9}

\title{Production of dark matter axions from collapse of string-wall systems}

\author{Takashi Hiramatsu}
\email{hiramatz@yukawa.kyoto-u.ac.jp}
\affiliation{Yukawa Institute for Theoretical Physics, Kyoto University, 
Kitashirakawa Oiwake-cho, Sakyo-ku, Kyoto 606-8502, Japan}
\author{Masahiro Kawasaki}
\email{kawasaki@icrr.u-tokyo.ac.jp}
\affiliation{Institute for Cosmic Ray Research, The University of Tokyo, 
5-1-5 Kashiwa-no-ha, Kashiwa City, Chiba 277-8582, Japan} 
\affiliation{Kavli Institute for the Physics and Mathematics of the Universe (WPI), Todai Institutes for Advanced Study, 
The University of Tokyo,
5-1-5 Kashiwa-no-ha, Kashiwa City, Chiba 277-8582, Japan}
\author{Ken'ichi Saikawa}
\email{saikawa@icrr.u-tokyo.ac.jp}
\affiliation{Institute for Cosmic Ray Research, The University of Tokyo, 
5-1-5 Kashiwa-no-ha, Kashiwa City, Chiba 277-8582, Japan}
\author{Toyokazu Sekiguchi}
\email{sekiguti@a.phys.nagoya-u.ac.jp}
\affiliation{Graduate School of Science, Nagoya University,
Furo-cho, Chikusa-ku, Nagoya City, Aichi 464-8602, Japan}
 
\date{\today}

\begin{abstract}
We analyze the spectrum of axions radiated from collapse of domain walls, which have received less attention in the literature.
The evolution of topological defects related to the axion models is investigated by performing field-theoretic lattice simulations.
We simulate the whole process of evolution of the defects, including the formation of global strings, the formation of domain walls and
the annihilation of the defects due to the tension of walls.
The spectrum of radiated axions has a peak at the low frequency, which implies that axions produced by the collapse of domain walls are not highly relativistic.
We revisit the relic abundance of cold dark matter axions and find that the contribution from the decay of defects can be comparable with
the contribution from strings. This result leads to a more severe upper bound on the axion decay constant.
\end{abstract}


\maketitle

\begin{widetext}
\tableofcontents\vspace{5mm}
\end{widetext}

\section{\label{sec1}Introduction}
The standard model of particle physics has been well established, except for several unresolved problems.
Among them, the strong CP problem in quantum chromodynamics (QCD) remains as a mystery.
The most attractive solution is the celebrated Peccei-Quinn (PQ) mechanism~\cite{1977PhRvL..38.1440P},
which introduces an global U(1) symmetry that has to be spontaneously broken at some high-energy scale.
This spontaneous breaking of the global symmetry predicts an existence of a (pseudo) Nambu-Goldstone boson, called the axion~\cite{1978PhRvL..40..223W}.
Soon after the proposal, it was found that the prototype model of the axion conflicts with experiments~\cite{1978PhRvD..18.1607D}, and it was ruled out.
However, it was argued that models with a higher symmetry breaking scale denoted as $F_a$ (the axion decay constant)
can still avoid the experimental constraints~\cite{1979PhRvL..43..103K,1981PhLB..104..199D}.
The essential point is that the couplings between axions and other fields are suppressed by a large factor of the symmetry breaking scale $\sim 1/F_a$.
These models are called ``invisible axions'' because of their smallness of coupling with matter.

This invisibleness leads to a cosmological consequence.
It turns out that almost stable coherently oscillating axion fields play a role of the cold dark matter filled in the universe~\cite{1983PhLB..120..127P}.
The present density of the cold dark matter is measured to be $\Omega_{\mathrm{CDM}}\approx 0.23$~\cite{2011ApJS..192...18K}.
Requiring that the present abundance of axions must not exceed the observed value, one obtains the upper bound for the symmetry breaking scale $F_a<10^{12}$GeV.
In other words, axions are good candidates of the dark matter if $F_a$ is as large as $10^{12}$GeV.

Furthermore, the astrophysical observations give a lower bound on $F_a$.
For example, from the helium-burning lifetimes of horizontal branch stars in the globular clusters one can obtain a bound $F_a> {\cal O}(1)\times 10^7$GeV~\cite{1999ARNPS..49..163R}.
Also, the cooling time of white-dwarfs~\cite{1986PhLB..166..402R} gives a bound
$F_a>{\cal O}(1)\times 10^8$GeV for the Dine-Fischler-Srednicki-Zhitnitsky (DFSZ) model~\cite{1981PhLB..104..199D}.
The most stringent bound comes from the energy loss rate of the supernova 1987A~\cite{1990PhR...198....1R}, which gives $F_a\gtrsim 10^9$GeV both for
the Kim-Shifman-Vainshtein-Zakharov (KSVZ) model~\cite{1979PhRvL..43..103K} and for the DFSZ model.
Combined with the cosmological bound described above, we can constrain the axion models into the parameter region with $F_a\simeq 10^9$-$10^{12}$GeV.
This is called ``the classical axion window.''

It was argued that this axion window can be further constrained.
Since the axion field stays in a vacuum manifold of U(1) after the spontaneous breaking of the PQ symmetry,
the linear topological objects called strings are formed in the early universe.
Davis~\cite{1986PhLB..180..225D} first recognized that axions emitted by these strings give an additional cosmological abundance and
found that this additional contribution dominates over the coherent oscillations, which gave a more severe upper bound on the axion decay constant.
This suggestion is supported by various analytical and numerical studies given
by different authors~\cite{1989NuPhB.324..167D,1994PhRvL..73.2954B,1996PhRvL..76.2203B}.
However, Harari and Sikivie~\cite{1987PhLB..195..361H} presented a different scenario which claims that the abundance of axions produced by strings do not exceed that of coherent oscillations,
and gave a weaker upper bound on $F_a$. The subsequent numerical studies provided by~\cite{1991NuPhB.363..247H} supported this weaker bound.

This controversy about the contribution from strings arises from different assumptions on the spectrum of radiated axions.
In Refs.~\cite{1986PhLB..180..225D,1989NuPhB.324..167D,1994PhRvL..73.2954B,1996PhRvL..76.2203B}
it is assumed that the typical wavelength of radiated axions is given by the curvature size of global strings which is comparable to the size of the horizon, $k\sim 2\pi/t$,
based on the result of~\cite{1985PhRvD..32.3172D} which claims that closed loops or bent stings oscillate many times before they lose most of their energy.
Let us call this case scenario A~\cite{2008LNP...741...19S}.
In this scenario, the contribution from axions produced by strings becomes greater than that from coherent oscillations by a factor of $\ln(t/\delta_s)\approx 69$, where $\delta_s$
is the width of strings. On the other hand, Refs.~\cite{1987PhLB..195..361H,1991NuPhB.363..247H} suggest that
the motion and decay of the global strings are more ``turbulent.'' Let us call this scenario B.
In this case, strings lose their energy in one oscillation time, quickly decaying into small pieces.
Therefore, whole scales between the largest scale $\sim t$ and the smallest scale $\sim\delta_s$ give the same contribution to the power spectrum of radiated axions,
$dE/d\ln k\sim \mathrm{const.}$ or $dE/dk\sim 1/k$. In this scenario, it turns out that the contribution from strings is comparable to that from coherent oscillations.
This discrepancy between the two scenarios might be resolved by the field-theoretic global simulations including the cosmic expansion performed in~\cite{1999PhRvL..82.4578Y,2011PhRvD..83l3531H}.
In~\cite{1999PhRvL..82.4578Y,2011PhRvD..83l3531H}, it was concluded that the power spectrum of radiated axions has a sharp peak around the horizon scale 
$k_{\rm phys}\sim 2\pi/t$, supporting scenario A.

However, it is not sufficient to just consider the string contribution.
It was found that these strings are attached to surface-like field configurations called domain walls
when the axion acquires a mass due to the nonperturbative effect of QCD~\cite{1982PhRvL..48.1156S}.
Lyth~\cite{1992PhLB..275..279L} pointed out that the annihilation of these domain walls produces additional radiation of axions.
Subsequently, authors in~\cite{1994PhRvD..50.4821N} and~\cite{1999PhRvD..59b3505C} investigated this process,
but conclusions of these studies are different from each other.
Nagasawa and Kawasaki~\cite{1994PhRvD..50.4821N} found that axions produced by the collapse of domain walls are mildly relativistic,
and this contribution can exceed that from strings. On the other hand, in the study given by Chang, Hagmann and Sikivie~\cite{1999PhRvD..59b3505C},
the mean energy of axions produced by the decay of domain walls was estimated to be larger than that obtained in~\cite{1994PhRvD..50.4821N} by a factor of 20.
This leads to the conclusion that axions produced by the collapse of walls are subdominant compared with that produced by strings.
The conclusion of~\cite{1999PhRvD..59b3505C} relies on the following reasoning.
Since domain walls are bounded by strings, the wall energy is converted into the kinetic energy of strings.
Then, if we assume that scenario B is correct, the spectrum of radiated axions becomes hard ($dE/dk\sim 1/k$).
However, as we described above, the recent network simulation of global strings supports scenario A.
Therefore it is not so clear whether the domain wall contribution is significant or not.

We point out that this discrepancy on the domain wall contribution is analogous to that on the axionic string contribution, and
again it can be resolved by preforming the full field-theoretic network simulations.
In this paper, we aim to determine the contribution of axions produced by the collapse of domain walls bounded by strings
and give the total relic abundance of cold axions including all production mechanisms (i.e. coherent oscillation, string decay, and wall decay).
We perform the three-dimensional lattice simulation of the scalar field and follow the whole processes relevant to the axion production
(from the PQ phase transition to the QCD phase transition).
This analysis gives a result with least theoretical uncertainties, in the sense that all field configurations are determined by a first principle.

We note that the above discussions are only applicable to the case in which PQ symmetry is broken after inflation.
If PQ symmetry is broken before inflation, there are no contributions from strings and domain walls
since the population of these defects is diluted in the inflationary era.
In this case, isocurvature fluctuations of the axion field gives some imprints on anisotropies of cosmic microwave background (CMB)
observed today~\cite{2008PThPh.120..995K}.
This observation gives severe constraints on axion models and requires significant amounts of fine-tunings in the model parameters (called ``the anthropic axion window'')~\cite{2011JCAP...07..021M}.
In this paper, we do not consider this possibility and simply assume that PQ symmetry is broken after inflation.

The organization of this paper is as follows.
In Sec.~\ref{sec2}, we explain the cosmological scenario in which the PQ symmetry breaking occurs after inflation and give a qualitative description of topological defects.
In Sec.~\ref{sec3}, we describe the analysis method which we used in the numerical studies.
We present the result of the numerical simulations in Sec.~\ref{sec4}.
Using the numerical results, we calculate the cosmological abundance of cold axions in Sec.~\ref{sec5}.
Finally, we conclude in Sec.~\ref{sec6}.

\section{\label{sec2}Cosmology with Peccei-Quinn field}
In this section, we give an overview of the cosmological aspects of the PQ mechanism, especially concentrating on the role of topological defects.
Depending on the model parameters, these defects can become either stable or unstable.
First we introduce theoretical basics and discuss some cosmological consequences.
We also give a comment on the time scale of the dynamics.
\subsection{\label{sec2-1}Formation of topological defects and their fates}
We will follow the cosmological evolution of a complex scalar field $\Phi$ (the Peccei-Quinn field) with the Lagrangian density
\begin{equation}
{\cal L} = \frac{1}{2}|\partial_{\mu}\Phi|^2- V_{\mathrm{eff}}[\Phi;T], \label{eq2-1-1}
\end{equation}
where $V_{\mathrm{eff}}[\Phi;T]$ is the finite-temperature effective potential for the scalar field.
At sufficiently high temperature ($T\gtrsim F_a$), we assume that $\Phi$ is in the thermal equilibrium, and the effective potential is given by
\begin{equation}
V_{\mathrm{eff}} [\Phi;T] = \frac{\lambda}{4}(|\Phi|^2-\eta^2)^2 + \frac{\lambda}{6}T^2|\Phi|^2,\label{eq2-1-2}
\end{equation}
where we neglect the couplings with other fields for simplicity.
Inspection of the form of the effective potential~(\ref{eq2-1-2}) indicates that the PQ phase transition occurs at the temperature
\begin{equation}
T = T_c \equiv \sqrt{3}\eta. \label{eq2-1-3}
\end{equation}
After that, the scalar field gets vacuum expectation value $|\langle\phi\rangle|^2=\eta^2$, and the U(1) symmetry is spontaneously broken.
Then, due to the spontaneous breaking of the U(1) symmetry, cosmic strings are formed.
Because of the causality, the population of these strings in the Hubble volume tends to remain in the value of ${\cal O}$(1) (the scaling solution)~\cite{1994csot.book.....V}.
In order to satisfy this scaling property, long strings lose their energy by emitting closed loops of strings.
These loops decay by radiating axion particles with the wavelength comparable to
the horizon size~\cite{1986PhLB..180..225D,1989NuPhB.324..167D,1999PhRvL..82.4578Y,2011PhRvD..83l3531H}.
The production of axions from decaying string loops continues until the time when the string networks disappear due to the mechanism which we describe below.

When the temperature of the universe becomes comparable to the QCD scale ($\Lambda\sim 100$MeV), the nonperturbative nature of QCD becomes relevant.
We can describe this effect by adding the following term in the effective potential~(\ref{eq2-1-2}).
\begin{equation}
V(\theta) = \frac{m_a^2\eta^2}{N_{\mathrm{DW}}^2}(1-\cos N_{\mathrm{DW}}\theta), \label{eq2-1-4}
\end{equation}
where $\theta$ is the phase direction of the complex scalar field (i.e. $\Phi=|\Phi|e^{i\theta}$), $N_{\mathrm{DW}}$ is an integer which is determined by considering the color anomaly~\cite{1982PhRvL..48.1156S}, and $m_a$ is the mass of the axion.
The mass of the axion depends on the temperature $T$, if the temperature is sufficiently high ($T\gtrsim \Lambda$).
Recently, Wantz and Shellard~\cite{2010PhRvD..82l3508W} presented the temperature dependence of $m_a$ which is valid at all temperatures
within the interacting instanton liquid model~\cite{2010NuPhB.829..110W}. Fitting the numerical result, they obtained the power-law expression for $m_a(T)$
\begin{equation}
m_a(T)^2 = c_T\frac{\Lambda^4}{F_a^2}\left(\frac{T}{\Lambda}\right)^{-n}, \label{eq2-1-5}
\end{equation}
where $n=6.68$, $c_T=1.68\times 10^{-7}$, and $\Lambda=400$MeV.
This power-law expression should be cut off by hand once it exceeds the zero-temperature value $m_a(T=0)$, where
\begin{equation}
m_a(0)^2 = c_0\frac{\Lambda^4}{F_a^2}, \label{eq2-1-6}
\end{equation}
and $c_0=1.46\times 10^{-3}$.

The existence of the QCD potential~(\ref{eq2-1-4}) explicitly breaks the original U(1) PQ symmetry down to its discrete subgroup $Z_{N_{\mathrm{DW}}}$,
in which the angular direction possesses the shift symmetry $\theta \to \theta + 2\pi k/N_{\mathrm{DW}}$ ($k=0,1,\dots,N_{\mathrm{DW}}-1$).
This $Z_{N_{\mathrm{DW}}}$ symmetry is also spontaneously broken because of the vacuum expectation value of the axion field.
As a consequence, $N_{\mathrm{DW}}$ domain walls attached to strings are formed~\cite{1982PhRvL..48.1156S}.
The structure of domain walls depends on the number $N_{\mathrm{DW}}$.
If $N_{\mathrm{DW}}>1$, these domain walls are stable and dominate the energy density of the universe,
which leads to the discrepancy with the cosmological observations and called the domain wall problem~\cite{1974JETP...40....1Z}.
On the other hand, if $N_{\mathrm{DW}}=1$, networks of domain walls are unstable, since the string is attached by only one domain wall.
Such a piece of the domain wall bounded by string can easily chop the larger one, or shrink itself due to the tension of the domain wall~\cite{1987NuPhB.283..591B}.
Hence the networks of domain walls bounded by strings disappear immediately after the formation.
In this case we can avoid the domain wall problem, and we assume $N_{\mathrm{DW}}=1$ in the rest of this paper.
Note that, it is possible to avoid the domain wall problem even in the case with $N_{\mathrm{DW}}>1$, by introducing the explicit $Z_{N_{\mathrm{DW}}}$
breaking term in the Lagrangian~\cite{1982PhRvL..48.1156S,1983PhRvD..27..332H}.
This kind of models also lead to interesting phenomenology~\cite{2011JCAP...08..030H}, and we will present the detailed analysis in another publication~\cite{HKSS2012}.

In the case with $N_{\mathrm{DW}}=1$, the annihilation of string-wall networks occurs around the time of the QCD phase transition.
At this time, potential energy stored in domain walls and strings is released as radiations of
axion particles~\cite{1992PhLB..275..279L,1994PhRvD..50.4821N,1999PhRvD..59b3505C}.
As we noted in Sec.~\ref{sec1}, our interest is to determine whether the population of axions produced by this mechanism is comparable or negligible
in comparison with other contributions such as axions produced by oscillating string loops and the coherent oscillation of the homogeneous field.
We will return to this issue in Sec.~\ref{sec5} after we present the result of the numerical study in Sec.~\ref{sec4}.
\subsection{\label{sec2-2}Typical time scale of the dynamics}
Before going to the numerical investigations, let us estimate the typical time scale of the annihilation process.
Since the tension of domain walls, which causes the decay of string-wall networks, is induced by the existence
of the axion mass, it is important to note the time at which the axion field begins to ``feel" the mass energy.
Let us denote this time as $t_1$, which is defined by the condition
\begin{equation}
m_a(T_1) = 3H(t_1), \label{eq2-2-1}
\end{equation}
where $T_1$ is the temperature at the time $t_1$, and $H(t_1)$ is the Hubble parameter at that time.
Using the temperature dependence of $m_a(T)$ given by Eq.~(\ref{eq2-1-5}), we find
\begin{equation}
T_1 = 0.981\mathrm{GeV}\left(\frac{g_{*,1}}{70}\right)^{-1/(4+n)}\left(\frac{F_a}{10^{12}\mathrm{GeV}}\right)^{-2/(4+n)}\left(\frac{\Lambda}{400\mathrm{MeV}}\right)
\quad(\mathrm{for}\quad T_1\gtrsim 103\mathrm{MeV}), \label{eq2-2-2}
\end{equation}
or
\begin{equation}
T_1 = 42.3\mathrm{GeV}\left(\frac{g_{*,1}}{70}\right)^{-1/4}\left(\frac{F_a}{10^{12}\mathrm{GeV}}\right)^{-1/2}\left(\frac{\Lambda}{400\mathrm{MeV}}\right)
\quad(\mathrm{for}\quad T_1\lesssim 103\mathrm{MeV}), \label{eq2-2-3}
\end{equation}
where $g_{*,1}$ is the radiation degree of freedom at the time $t_1$.
Equation~(\ref{eq2-2-2}) is valid only for $T_1\gtrsim103$MeV, which corresponds
to the case in which the condition given by Eq.~(\ref{eq2-2-1}) is satisfied before $m_a(T)$ becomes the 
zero-temperature value $m_a(0)$. We must use another expression (\ref{eq2-2-3}) if $T_1<103$MeV. However,
if we fix the values as $g_{*,1}=70$ and $\Lambda=400$MeV, this turnover occurs around the value $F_a\simeq1.7\times10^{17}$GeV.
Therefore, we can simply use Eq.~(\ref{eq2-2-2}) as long as we assume that $F_a<1.7\times10^{17}$.
The temperature given by Eq.~(\ref{eq2-2-2}) or Eq.~(\ref{eq2-2-3}) corresponds to the time
\begin{equation}
t_1 = 3.01\times 10^{-7}\mathrm{sec} \left(\frac{g_{*,1}}{70}\right)^{-n/2(4+n)}\left(\frac{F_a}{10^{12}\mathrm{GeV}}\right)^{4/(4+n)}\left(\frac{\Lambda}{400\mathrm{MeV}}\right)^{-2}. \quad(\mathrm{for}\quad T_1\gtrsim 103\mathrm{MeV}), \label{eq2-2-4}
\end{equation}
or
\begin{equation}
t_1 = 1.61\times 10^{-10}\mathrm{sec}\left(\frac{F_a}{10^{12}\mathrm{GeV}}\right)\left(\frac{\Lambda}{400\mathrm{MeV}}\right)^{-2}. \quad(\mathrm{for}\quad T_1\lesssim 103\mathrm{MeV}). \label{eq2-2-5}
\end{equation}

Another relevant time scale is the time when the string-wall networks annihilate themselves.
One might guess that this occurs when the tension of domain walls dominates over that of strings.
We denote this time as $t_2$, which is defined by
\begin{equation}
\sigma_{\mathrm{wall}}(t_2) = \mu_{\mathrm{str}}(t_2)/t_2, \label{eq2-2-6}
\end{equation}
where $\sigma_{\mathrm{wall}}=9.23m_a(T)F_a^2$ is the surface mass density of domain walls~\cite{1985PhRvD..32.1560H},
$\mu_{\mathrm{str}}=\pi F_a^2\ln\left(\frac{t/\sqrt{\xi}}{\delta_s}\right)$ is the mass energy of the strings per unit length,
$\xi$ is the length parameter of strings defined by Eq.~(\ref{eq4-7}), and $\delta_s=1/\sqrt{\lambda}\eta$ is the width of the core of strings.
From Eq.~(\ref{eq2-2-6}), we obtain
\begin{equation}
t_2 = 8.43\times10^{-5}\mathrm{sec}\left(\frac{g_{*,2}}{70}\right)^{-n/2(n+4)}\left(\frac{F_a}{10^{12}\mathrm{GeV}}\right)^{4/(n+4)}\left(\frac{\Lambda}{400\mathrm{MeV}}\right)^{-2}
\quad(\mathrm{for}\quad T_2\gtrsim 103\mathrm{MeV}), \label{eq2-2-7}
\end{equation}
or
\begin{equation}
t_2 = 2.53\times10^{-9}\mathrm{sec}\left(\frac{F_a}{10^{12}\mathrm{GeV}}\right)\left(\frac{\Lambda}{400\mathrm{MeV}}\right)^{-2}
\quad(\mathrm{for}\quad T_2\lesssim 103\mathrm{MeV}), \label{eq2-2-8}
\end{equation}
and the corresponding temperature
\begin{equation}
T_2 = 0.586\mathrm{GeV}\left(\frac{g_{*,2}}{70}\right)^{-1/(n+4)}\left(\frac{F_a}{10^{12}\mathrm{GeV}}\right)^{-2/(n+4)}\left(\frac{\Lambda}{400\mathrm{MeV}}\right)\quad(\mathrm{for}\quad T_2\gtrsim 103\mathrm{MeV}), \label{eq2-2-9}
\end{equation}
or
\begin{equation}
T_2 = 10.7\mathrm{GeV}\left(\frac{g_{*,2}}{70}\right)^{-1/4}\left(\frac{F_a}{10^{12}\mathrm{GeV}}\right)^{-1/2}\left(\frac{\Lambda}{400\mathrm{MeV}}\right)\quad(\mathrm{for}\quad T_2\lesssim 103\mathrm{MeV}), \label{eq2-2-10}
\end{equation}
where $g_{*,2}$ is the radiation degree of freedom at the time $t_2$, and we substituted the typical value $\ln\left(\frac{t/\sqrt{\xi}}{\delta_s}\right)\approx 69$.
The ratio $t_2/t_1\simeq 3$ indicates that
it takes a few Hubble times for the mass term to become effective.

To summarize, the history of the universe in the case with $N_{\mathrm{DW}}=1$ is described as follows.
First, the PQ phase transition occurs at $T\simeq T_c$, and global strings are formed.
Next, QCD effects become relevant at $T=T_1$, and domain walls are attached to strings.
Finally, these domain walls bounded by strings disappear around the temperature $T=T_2$.
We will see that these processes actually occur in the field-theoretic lattice simulations in Sec.~\ref{sec4}.
\section{\label{sec3}Analysis method}
In this section, we describe the method to calculate the spectrum of axions produced by the decay of string-wall networks.
Our aim is to extract the pure component of the axion field produced by collapse of the networks, from simulated data of the scalar field $\Phi$.
In general, the data of $\Phi$ contain other components, which can be enumerated as follows:
\begin{enumerate}
\item {\it Initial fluctuations}. In numerical simulations, we give the initial conditions as Gaussian random fluctuations [see Eqs.~(\ref{eq3-1-21}),~(\ref{eq3-1-22}) and~(\ref{eq3-1-23})].
These fluctuations are diluted away by the cosmic expansion, but might not be completely negligible even at the final time of the simulation,
since the dynamical range of the numerical simulation is short. Therefore, they can contaminate the final form of the spectrum of radiated axions.
\item {\it Radiations from strings}. As we mentioned in Sec.~\ref{sec2-1}, oscillating loops of strings radiate axions during the time between
the string formation ($T=T_c$) and the domain wall formation ($T=T_1$).
This contribution must be distinguished from the wall-decay contribution which is produced {\it after} the time $t_1$.
\item {\it Core of defects}. In the core of strings, the energy density of the scalar field is higher than that of free axions.
This can be regarded as another contamination on the spectrum of radiated axions~\cite{2011PhRvD..83l3531H}.
\end{enumerate}

\begin{figure}[htbp]
\begin{center}
\includegraphics[scale=0.65]{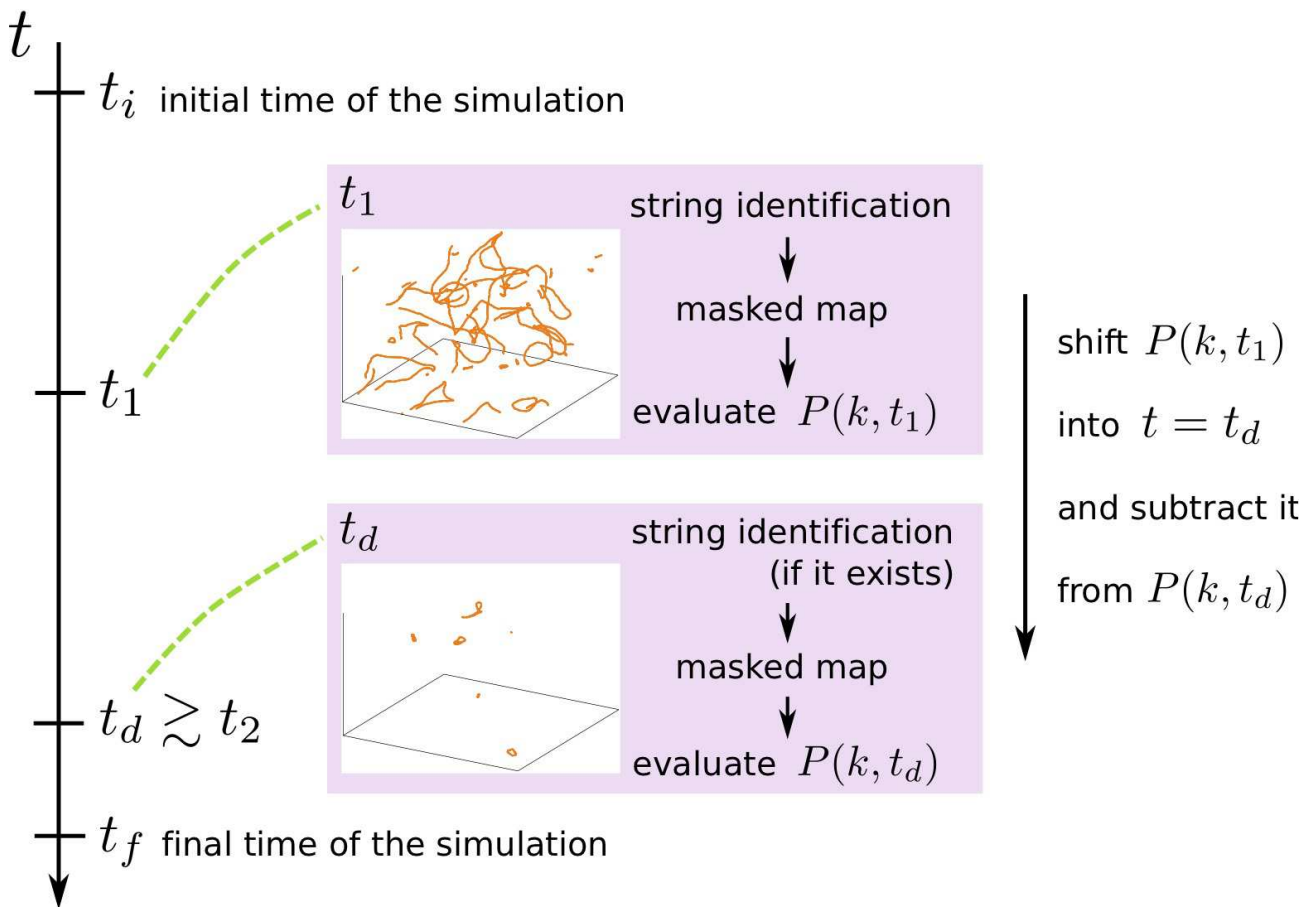}
\end{center}
\caption{Schematics of the procedure to estimate the power spectrum of radiated axions.}
\label{fig1}
\end{figure}

Figure~\ref{fig1} shows the pipeline of removing these contaminations.
To remove the contaminations from the core of strings, we mask the region near the position of the core of strings
and estimate the power spectrum which contains only the contribution from free radiations.
We calculate the power spectrum in two time slices, the time at which the mass of the axion becomes relevant ($t=t_1$)
and the time at which the decay of string-wall networks completes ($t=t_d$).
Then, we subtract the spectrum evaluated at $t_1$ from that evaluated at $t_d$ in order to remove the contributions which come from
initial fluctuations and radiations from strings.
We will give a more detailed description of these procedures in the following subsections.
First, we establish the formulations for field-theoretic simulations and the notations to present the result of the numerical study in Sec.~\ref{sec3-1}.
Then, we describe how to calculate the power spectrum of radiated axions in Secs.~\ref{sec3-2} and \ref{sec3-3}.
Finally, in Sec.~\ref{sec3-4} we comment on the subtraction of other radiation components.

\subsection{\label{sec3-1}Formulations}
We assume the flat Friedmann-Robertson-Walker background in which the line element is given by
\begin{equation}
ds^2 = dt^2-R^2(t)\delta_{ij}dx^idx^j, \label{eq3-1-1}
\end{equation}
where $R(t)$ is the scale factor of the universe. The equation of motion for $\Phi$ is obtained by varying the Lagrangian given by Eq.~(\ref{eq2-1-1})
with the effective potential given by
\begin{equation}
V_{\mathrm{eff}}[\Phi;T] = \frac{\lambda}{4}(|\Phi|^2-\eta^2)^2 + \frac{\lambda}{6}T^2|\Phi|^2 + m_a(T)^2\eta^2\left(1-\frac{|\Phi|}{\eta}\cos\theta\right). \label{eq3-1-2}
\end{equation}
Note that the last term of Eq.~(\ref{eq3-1-2}) is different from the QCD potential~(\ref{eq2-1-4}).
We find that the simulation becomes unstable when using the potential given by Eq.~(\ref{eq2-1-4}) since this potential is not well defined at $|\Phi|=0$.
The modified potential given by Eq.~(\ref{eq3-1-2}) avoids this singularity since there is a factor $|\Phi|$ in front of the cosine term.
The difference between Eq.~(\ref{eq2-1-4}) and Eq.~(\ref{eq3-1-2}) is not important in the bulk region on which $|\Phi|=\eta$,
and we observe that the quantitative behavior of topological defects such as time evolution of the length of strings is unchanged, except the existence of the numerical instability.

We decompose the complex scalar field into its real and imaginary part, such that
\begin{equation}
\Phi = \phi_1 + i\phi_2, \label{eq3-1-3}
\end{equation}
where $\phi_1$ and $\phi_2$ are real variables. The equations of motion for two real scalar fields $\phi_1$ and $\phi_2$ are given by,
\begin{align}
\ddot{\phi}_1 + 3H\dot{\phi}_1 - \frac{\nabla^2}{R^2}\phi_1 =& -\lambda\phi_1(|\Phi|^2-\eta^2) - \frac{\lambda}{3}T^2\phi_1 + m_a^2(T)\eta, \label{eq3-1-4}\\
\ddot{\phi}_2 + 3H\dot{\phi}_2 - \frac{\nabla^2}{R^2}\phi_2 =& -\lambda\phi_2(|\Phi|^2-\eta^2) - \frac{\lambda}{3}T^2\phi_2. \label{eq3-1-5}
\end{align}
If we use the conformal time $d\tau\equiv dt/R$ rather than the cosmic time, these equations can be rewritten in the following form.
\begin{align}
\bar{\phi}_1'' - \nabla^2\bar{\phi}_1 =& -\lambda\bar{\phi}_1(|\bar{\Phi}|^2-R^2\eta^2) - \frac{\lambda}{3}R^2T^2\bar{\phi}_1 + R^3m_a^2(T)\eta, \label{eq3-1-6}\\
\bar{\phi}_2'' - \nabla^2\bar{\phi}_2 =& -\lambda\bar{\phi}_2(|\bar{\Phi}|^2-R^2\eta^2) - \frac{\lambda}{3}R^2T^2\bar{\phi}_2, \label{eq3-1-7}
\end{align}
where the prime represents a derivative with respect to $\tau$, and we introduced rescaled variable
\begin{equation}
\bar{\Phi} \equiv R\Phi. \label{eq3-1-8}
\end{equation}
Note that, in deriving Eqs.~(\ref{eq3-1-6}) and (\ref{eq3-1-7}), we assume a radiation dominated background, in which $R''=0$.

In the radiation-dominated universe, the time and temperature are related by the Friedmann equation
\begin{equation}
\frac{1}{4t^2} = H^2 = \frac{8\pi G}{3}\frac{\pi^2}{30}g_*T^4, \label{eq3-1-9}
\end{equation}
where $G$ is Newton's gravitational constant and $g_*$ is the relativistic degree of freedom.
For convenience in the numerical study, we introduce a dimensionless quantity
\begin{equation}
\zeta \equiv \sqrt{\frac{45}{16\pi^3Gg_*}}\frac{1}{\eta}. \label{eq3-1-10}
\end{equation}
Using this parameter, Eq.~(\ref{eq3-1-9}) can be written as
\begin{equation}
t = \frac{\zeta\eta}{T^2}. \label{eq3-1-11}
\end{equation}

In the following, we normalize all the dimensionful quantities in the unit of $\tau_c$, which is the conformal time at which PQ phase transition occurs [cf. Eq.~(\ref{eq2-1-3})],
\begin{equation}
\Phi \to \Phi\tau_c,\qquad T \to T\tau_c, \qquad x \to x/\tau_c, \quad\mathrm{etc}. \label{eq3-1-12}
\end{equation}
Also, we introduce the normalized initial Hubble parameter as an input parameter of the numerical simulation
\begin{equation}
H(t=t_i) \to \tau_cH(t=t_i) \equiv \alpha, \label{eq3-1-13}
\end{equation}
and we set the scaling parameter at the initial time into unity
\begin{equation}
R(t_i) = 1. \label{eq3-1-14}
\end{equation}
Note the following relations:
\begin{equation}
R(\tau_c) = \tau_c/\tau_i = \alpha \qquad \mathrm{and}\qquad\tau_c = \frac{2\zeta}{3\eta\alpha}. \label{eq3-1-15}
\end{equation}
By using Eq.~(\ref{eq3-1-15}), we can enumerate the various relations in the unit (\ref{eq3-1-12}),
\begin{equation}
\tau_i = 1/\alpha, \quad
R(\tau) = \alpha\tau, \quad
T_i = \frac{2\zeta}{\sqrt{3}} \quad \mathrm{and} \quad
\eta = \frac{2\zeta}{3\alpha}. \label{eq3-1-16}
\end{equation}
Then, equations of motion for scalar fields (\ref{eq3-1-6}) and (\ref{eq3-1-7}) reduce to
\begin{align}
\bar{\phi}_1'' - \nabla^2\bar{\phi}_1 =& -\lambda\bar{\phi}_1\left(|\bar{\Phi}|^2 - \frac{4\tau^2\zeta^2}{9}\right) - \frac{4\lambda}{9}\zeta^2\bar{\phi}_1 + \frac{8\tau^3\zeta^3}{27}\left(\frac{m_a(T)}{\eta}\right)^2, \label{eq3-1-17}\\
\bar{\phi}_2'' - \nabla^2\bar{\phi}_2 =& -\lambda\bar{\phi}_2\left(|\bar{\Phi}|^2 - \frac{4\tau^2\zeta^2}{9}\right) - \frac{4\lambda}{9}\zeta^2\bar{\phi}_2. \label{eq3-1-18}
\end{align}
Here, the ratio between the axion mass~(\ref{eq2-1-5}) and the axion decay constant can be written as
\begin{equation}
m_a(T)^2/\eta^2 = c_T\kappa^{n+4}\left(\frac{T}{F_a}\right)^{-n}, \label{eq3-1-19}
\end{equation}
where $\kappa$ is the ratio between the QCD scale and the PQ scale,
\begin{equation}
\kappa\equiv \Lambda/F_a = \Lambda/\eta. \label{eq3-1-20}
\end{equation}

Assuming that $\Phi$ is in the thermal equilibrium at the high temperature,
we give the initial conditions such that the two real scalar fields satisfy the renormalized correlation function
\begin{align}
\langle\phi_a({\bf x})\phi_b({\bf y})\rangle &= \delta_{ab}\int\frac{d^3k}{(2\pi)^3}\frac{n_k}{\omega_k}e^{i{\bf k\cdot(x-y)}}, \label{eq3-1-21} \\
\langle\dot{\phi}_a({\bf x})\dot{\phi}_b({\bf y})\rangle &= \delta_{ab}\int\frac{d^3k}{(2\pi)^3}\omega_kn_ke^{i{\bf k\cdot(x-y)}}, \label{eq3-1-22}\\
\langle\phi_a({\bf x})\dot{\phi}_b({\bf y})\rangle &= 0, \label{eq3-1-23}
\end{align}
where
\begin{equation}
n_k = \frac{1}{e^{\omega_k/T_i}-1},\qquad \omega_k = \sqrt{k^2+m_{\mathrm{eff}}^2}, \label{eq3-1-24}
\end{equation}
and $m_{\mathrm{eff}}^2 \equiv \partial^2V_{\mathrm{eff}}/\partial\Phi^*\partial\Phi|_{\Phi=0}$ is the effective mass of the scalar fields at the initial time.
Note that, in Eqs.~(\ref{eq3-1-21}) and (\ref{eq3-1-22}), we subtracted the vacuum fluctuations which contribute as a divergent term when we perform the integral of $k$.

In the momentum space, these correlation functions can be written as
\begin{align}
\langle\phi_a({\bf k})\phi_b({\bf k'})\rangle = \delta_{ab}\frac{n_k}{\omega_k}(2\pi)^3\delta^{(3)}({\bf k+k'}), \label{eq3-1-25} \\
\langle\dot{\phi}_a({\bf k})\dot{\phi}_b({\bf k'})\rangle = \delta_{ab}\omega_kn_k(2\pi)^3\delta^{(3)}({\bf k+k'}), \label{eq3-1-26}
\end{align}
where $\phi_a({\bf k})$ is the Fourier transform of $\phi({\bf x})$. Since $\phi_a({\bf k})$ and $\dot{\phi}_a({\bf k})$
are uncorrelated in the momentum space, we generate $\phi_a({\bf k})$ and $\dot{\phi}_a({\bf k})$ in the momentum space randomly
following the Gaussian distribution with
\begin{eqnarray}
&\langle|\phi_a({\bf k})|^2\rangle = \frac{n_k}{\omega_k}V, \qquad \langle|\dot{\phi}_a({\bf k})|^2\rangle = n_k\omega_kV,& \label{eq3-1-27}\\
&\mathrm{and}\qquad \langle\phi_a({\bf k})\rangle = \langle\dot{\phi}_a({\bf k})\rangle = 0,& \label{eq3-1-28}
\end{eqnarray}
for each $a=1$ and $2$. Then we transform them into the configuration space and obtain the initial field configurations $\phi({\bf x})$ and $\dot{\phi}({\bf x})$.
Here we used $(2\pi)^3\delta^{(3)}(0)\simeq V$, where $V=L^3$ is the comoving volume of the simulation box and $L$ is the size of the simulation box (in the unit of $\tau_c$).

We solve the classical equations of motion given by Eqs.~(\ref{eq3-1-17}) and (\ref{eq3-1-18}) in the three-dimensional lattice with $512^3$ points.
We impose the periodic boundary condition in the simulation box.
The lattice code which we use in this paper is developed by combining the numerical codes used in~\cite{2011PhRvD..83l3531H} and~\cite{2011JCAP...08..030H}.
We use the fourth-order symplectic integration scheme~\cite{1990PhLA..150..262Y} to solve the time evolution of the fields.
The spatial derivative of the fields is evaluated by using the fourth-order finite-difference method.
\subsection{\label{sec3-2}Energy spectrum of axions}
We calculate the power spectrum of axion radiations $P(k,t)$ defined by
\begin{equation}
\frac{1}{2}\langle\dot{a}(t,{\bf k})^*\dot{a}(t,{\bf k'})\rangle = \frac{(2\pi)^3}{k^2}\delta^{(3)}({\bf k}-{\bf k}')P(k,t), \label{eq3-2-1}
\end{equation}
where $\langle\dots\rangle$ represents an ensemble average and $\dot{a}(t,{\bf k})$ is the Fourier component of the time derivative of the axion field
\begin{equation}
\dot{a}(t,{\bf k}) = \int d^3{\bf x}e^{i{\bf k\cdot x}}\dot{a}(t,{\bf x}). \label{eq3-2-2}
\end{equation}
The value of $\dot{a}(t,{\bf x})$ can be obtained from the simulated data of $\Phi$ and $\dot{\Phi}$
\begin{equation}
\dot{a}(t,{\bf x}) = \mathrm{Im}\left[\frac{\dot{\Phi}}{\Phi}(t,{\bf x})\right]. \label{eq3-2-3}
\end{equation}

The averaged kinetic energy of axions can be written as
\begin{equation}
\rho_{a,\mathrm{kin}}(t) = \left\langle\frac{1}{2}\dot{a}(t,{\bf x})^2\right\rangle = \int\frac{dk}{2\pi^2}P(k,t). \label{eq3-2-4}
\end{equation}
On the other hand, the total energy density of axions is given by
\begin{equation}
\rho_{a,\mathrm{tot}}(t) = \rho_{a,\mathrm{kin}}(t) +  \rho_{a,\mathrm{grad}}(t) +  \rho_{a,\mathrm{mass}}(t), \label{eq3-2-5}
\end{equation}
where $\rho_{a,\mathrm{grad}}(t)$ is the averaged gradient energy of axions and $\rho_{a,\mathrm{mass}}(t)$ is the averaged mass energy of axions
\begin{equation}
\rho_{a,\mathrm{grad}}(t) = \left\langle\frac{1}{2}|\nabla a(t,{\bf x})|^2\right\rangle, \qquad \rho_{a,\mathrm{mass}}(t) = \left\langle\frac{1}{2}m_a^2a(t,{\bf x})^2\right\rangle. \label{eq3-2-6}
\end{equation}
One can easily show that, if $a(t,{\bf x})$ is a free field,
\begin{equation}
\rho_{a,\mathrm{kin}}(t) = \rho_{a,\mathrm{grad}}(t) + \rho_{a,\mathrm{mass}}(t). \label{eq3-2-7}
\end{equation}
Therefore, $P(k,t)$ can be regarded as the energy spectrum of axions
\begin{equation}
\rho_{a,\mathrm{tot}}(t) = 2\rho_{a,\mathrm{kin}}(t) = 2\int\frac{dk}{2\pi^2}P(k,t). \label{eq3-2-8}
\end{equation}
\subsection{\label{sec3-3}Pseudo-power spectrum estimator}
If strings exist, the data of $\dot{a}(t,{\bf x})$ obtained by numerical simulations contain field values around moving strings, 
\begin{equation}
\dot{a}(t,{\bf x}) = \dot{a}_{\mathrm{free}}(t,{\bf x}) + (\mathrm{contamination\ from\ strings}), \label{eq3-3-1}
\end{equation}
where $\dot{a}_{\mathrm{free}}(t,{\bf x})$ is the contribution from free axion radiations.
This moving-string contribution can contaminate the spectrum of the axion radiations.
In order to subtract the contamination from strings we use the pseudo-power spectrum estimator (PPSE) introduced in \cite{2011PhRvD..83l3531H}.

We mask the contribution from the axion field near strings by introducing a window function
\begin{equation}
W({\bf x}) = 
\left\{
\begin{array}{l l}
0 & (\mathrm{near\ strings}) \\
1 & (\mathrm{elsewhere}) \\
\end{array}
\right.
. \label{eq3-3-2}
\end{equation}
Then, we obtain the masked axion field
\begin{equation}
\tilde{\dot{a}}({\bf x}) \equiv W({\bf x})\dot{a}({\bf x}) = W({\bf x})\dot{a}_{\mathrm{free}}({\bf x}), \label{eq3-3-3}
\end{equation}
or, in the Fourier space,
\begin{equation}
\tilde{\dot{a}}({\bf k}) = \int\frac{d^3{\bf k}'}{(2\pi)^3}W({\bf k}-{\bf k}')\dot{a}({\bf k}'). \label{eq3-3-4}
\end{equation}
We can compute the power spectrum by using the masked field in a simulation box,
\begin{equation}
\tilde{P}(k) \equiv \frac{k^2}{V}\int\frac{d\Omega_k}{4\pi}\frac{1}{2}|\tilde{\dot{a}}({\bf k})|^2, \label{eq3-3-5}
\end{equation}
where $V$ is the comoving volume of the simulation box and $\Omega_k$ is a unit vector representing the direction of ${\bf k}$.
However, this masked spectrum is not equivalent to the spectrum of radiated axions, $\langle\tilde{P}(k)\rangle\ne P_{\mathrm{free}}$,
where $P_{\mathrm{free}}(k)$ is defined by
\begin{equation}
\frac{1}{2}\langle\dot{a}_{\mathrm{free}}(t,{\bf k})^*\dot{a}_{\mathrm{free}}(t,{\bf k'})\rangle = \frac{(2\pi)^3}{k^2}\delta^{(3)}({\bf k}-{\bf k}')P_{\mathrm{free}}(k,t). \label{eq3-3-6}
\end{equation}
We can resolve this discrepancy by introducing a window weight matrix,
\begin{equation}
M(k,k') \equiv \frac{1}{V^2}\int\frac{d\Omega_k}{4\pi}\frac{d\Omega_{k'}}{4\pi}|W({\bf k} - {\bf k}')|^2, \label{eq3-3-7}
\end{equation}
and defining the PPSE of $P_{\mathrm{free}}(k)$,
\begin{equation}
P_{\mathrm{PPSE}}(k) \equiv \frac{k^2}{V}\int\frac{dk'}{2\pi^2}M^{-1}(k,k')\tilde{P}(k'), \label{eq3-3-8}
\end{equation}
with $M^{-1}(k,k')$ satisfying
\begin{equation}
\int\frac{k'^2dk'}{2\pi^2}M^{-1}(k,k')M(k',k'') = \frac{2\pi^2}{k^2}\delta(k-k''). \label{eq3-3-9}
\end{equation}
Then, we see that $\langle P_{\mathrm{PPSE}}(k)\rangle = P_{\mathrm{free}}(k)$ \cite{2011PhRvD..83l3531H}.

In the numerical simulations, first we calculate the masked power spectrum $\tilde{P}(k)$ and the matrix
$M(k,k')$ by using the data of $\Phi({\bf x})$ and $\dot{\Phi}({\bf x})$, then, we compute the power spectrum of free axions by using Eq.~(\ref{eq3-3-8}).

For the identification of strings, we use the method developed by~\cite{2011PhRvD..83l3531H}.
At each lattice segment surrounded by four neighboring grids in the simulation box, 
we identify the existence of the string by the condition $\Delta\theta > \pi$, where $\Delta\theta$ is the minimal phase interval occupied
by four grid points in the field space. We compute the position of string as a cross-point of two lines with $\phi_1=0$ and $\phi_2=0$ in the quadrate.
See~\cite{2011PhRvD..83l3531H} for details.
\subsection{\label{sec3-4}Subtraction of preexisting radiations}
In order to extract the spectrum of axions radiated from the decay of string-wall networks, we must subtract the
contributions of preexisting radiations, which have been created before the decay of domain walls, from the whole spectrum
calculated by using the field data obtained after the decay of networks.
The radiations created before the decay of domain walls are diluted due to the cosmic expansion,
and we evaluate this redshift factor before we perform the subtraction.
Note that the axion mass becomes non-negligible around the time of the decay of domain walls.
Hence, we cannot subtract the spectrum simply assuming that the spectrum is
diluted as $R^{-4}$, which is only applicable to massless particles.

From Eq.~(\ref{eq3-2-8}), the total energy density of axions can be written as
\begin{equation}
\rho_a(t) = \int\frac{d^3{\bf k}}{(2\pi)^3}\omega_a(k,t)n_a(k,t), \label{eq3-4-1}
\end{equation}
where
\begin{equation}
\omega_a(k,t) = \sqrt{m_a^2+k^2/R(t)^2} \label{eq3-4-2}
\end{equation}
is the energy of axions with momentum $k/R(t)$, and we define
\begin{equation}
n_a(k,t) \equiv 2\frac{P(k,t)}{\omega_a(k,t)k^2}. \label{eq3-4-3}
\end{equation}
We can regard $n_a(k,t)d^3{\bf k}/(2\pi)^3$ as the number density of axions which have comoving momentum within the range from $k$ to $k+dk$.
Therefore, we expect that $n_a(k,t)$ scales as $R(t)^{-3}$, if there are no changes in the number of axions.

Let us denote the time at which domain walls sufficiently decay as $t_d$
(this definition of $t_d$ may contain an ambiguity, which we discuss in the next section).
By using the fact that $n_a(k,t)\propto R(t)^{-3}$ if there is no absorption or production of axions,
and Eq.~(\ref{eq3-4-3}), we find the form of the spectrum 
of preexisting radiations produced before $t_1$, where $t_1$ is defined by Eq.~(\ref{eq2-2-1}), at the time $t_d$,
\begin{equation}
P_{\mathrm{rad}}(k,t_d) = P(k,t_1)\frac{\omega_a(k,t_d)}{\omega_a(k,t_1)}\left(\frac{R(t_1)}{R(t_d)}\right)^3, \label{eq3-4-4}
\end{equation}
where $P(k,t_1)$ is the spectrum evaluated at $t_1$. Subtracting the contribution $P_{\mathrm{rad}}(k,t_d)$
from the whole spectrum $P(k,t_d)$ evaluated at $t_d$, we obtain the spectrum of radiations produced after the
decay of domain walls
\begin{equation}
P_{\mathrm{dec}}(k,t_d) = P(k,t_d) - P_{\mathrm{rad}}(k,t_d). \label{eq3-4-5}
\end{equation}
\section{\label{sec4}Results of numerical simulations}
In numerical simulations, we can vary four parameters $\lambda$, $\kappa$, $\zeta$ and $\alpha$.
We set $\alpha=2.0$ which corresponds to the fact that $\tau_i=0.5\tau_c$.
Also, we choose the value of $\zeta$ as 3.0, which corresponds to the conditions $\eta=1.23\times 10^{17}$GeV and $g_*=100$.
It seems that this value of $\eta$ may be too high and affect the small scale dynamics of the system. We will discuss this point in the end of this section.
Note that, from Eq.~(\ref{eq3-1-16}) we see that $\eta=1$ in the unit of $\tau_c^{-1}$.
Other parameters that we used are summarized in Table~\ref{tab1}. The dynamical range of the simulation is estimated as $\tau_f/\tau_i = 24$.

\begin{table}[h]
\begin{center}
\caption{Parameters used in numerical simulations.}
\vspace{3mm}
\begin{tabular}{c c}
\hline\hline
Grid size ($N$) & 512 \\
Box size ($L$) & 20 \\
Total number of steps & 1150 \\
Time interval ($d\tau$) & 0.01 \\
$\lambda$ & 1.0 \\
$\kappa$ [Eq.~(\ref{eq3-1-20})] & varying \\
$\zeta$ [Eq.~(\ref{eq3-1-10})] & 3.0 \\
$\alpha$ [Eq.~(\ref{eq3-1-13})] & 2.0 \\
$c_T$ [Eq.~(\ref{eq2-1-5})] & 6.26 \\
$c_0$ [Eq.~(\ref{eq2-1-6})] & 1.0 \\
initial time ($\tau_i$) & 0.5 \\
final time ($\tau_f$) & 12.0 \\
\hline\hline
\label{tab1}
\end{tabular}
\end{center}
\end{table}

We must keep the following conditions,
in order to simulate the dynamics of the topological defects correctly:
\begin{itemize}
\item The width of global strings $\delta_s = 1/\eta\sqrt{\lambda}$ must be greater than the physical lattice spacing $\delta x_{\mathrm{phys}} = R(t)L/N$,
where $N=512$ is the number of grids, in order to maintain the resolution of the width of strings.
\item The Hubble radius $H^{-1}$ must be smaller than the box size $R(t)L$, to avoid the unphysical effect caused by the finiteness nature of the simulation box.
\end{itemize}
The physical scale of the Hubble radius and the width of strings divided by the physical lattice spacing are respectively
\begin{equation}
\frac{H^{-1}}{\delta x_{\mathrm{phys}}} = \frac{N}{L}\tau, \qquad \mathrm{and} \qquad \frac{\delta_s}{\delta x_{\mathrm{phys}}} = \frac{3N}{2L\zeta\tau\sqrt{\lambda}}. \label{eq4-1}
\end{equation}
For the parameters given in Table~\ref{tab1}, we get $H^{-1}/\delta x_{\mathrm{phys}}\simeq 307$ and $\delta_s/\delta x_{\mathrm{phys}}\simeq 1.07$ at the end of the simulation $\tau=\tau_f$.
Therefore, the conditions described above are satisfied even at the end of the simulation.

We also treat $c_T$ and $c_0$ defined in Eqs.~(\ref{eq2-1-5}) and (\ref{eq2-1-6}) as free parameters in numerical simulations.
In terms of these parameters, the time at which the value of $m_a(T)$ reaches the zero-temperature value $m_a(0)$ is written as
\begin{equation}
\tau_a = 1.73\times\left(\frac{c_0}{c_T}\right)^{1/n}\kappa^{-1}. \label{eq4-2}
\end{equation}
Here we used the conformal time with the unit of $\tau_c=1$. Also, the time $t_1$ given by Eq.~(\ref{eq2-2-1}) is rewritten as
\begin{equation}
\tau_1 = 1.52\times\left(\frac{3.0}{\zeta}\right)^{2/(4+n)}c_T^{-1/(n+4)}\kappa^{-1}. \label{eq4-3}
\end{equation}
Choosing the values of three parameters $c_0$, $c_T$ and $\kappa$ corresponds to the fact that
we tune the values of $\tau_a$, $\tau_1$, and $m_a(0)$ in the numerical simulations.
Unfortunately, due to the limitation of the dynamical range, we cannot choose all parameters to be realistic values.
One possible choice is to fix the ratio between $\tau_1$ and $\tau_a$ so that
\begin{equation}
\tau_1/\tau_a = 0.88\times\left(\frac{c_T}{c_0}\right)^{1/n}c_T^{-1/(n+4)}\left(\frac{3.0}{\zeta}\right)^{2/(n+4)} = 0.97\times\left(\frac{3.0}{\zeta}\right)^{2/(n+4)}, \label{eq4-4}
\end{equation}
where the last equality follows from the realistic values $c_T=1.68\times10^{-7}$ and $c_0=1.46\times10^{-3}$.
In order to satisfy the condition~(\ref{eq4-4}), we must choose $c_T$ and $c_0$ so that
\begin{equation}
c_Tc_0^{-(n+4)/4} \simeq 6.26. \label{eq4-5}
\end{equation}
In the numerical simulations, we use the values $c_0=1.0$ and $c_T=6.26$ which satisfy the above condition.
In this case the expression for $\tau_2$ becomes
\begin{equation}
\tau_2 = 1.26\times \left(\frac{3.0}{\zeta}\right)^{2/(4+n)}\left(\frac{\beta}{4}\right)^{2/(4+n)}\left(\frac{6.26}{c_T}\right)^{1/(4+n)}\kappa^{-1}, \label{eq4-6}
\end{equation}
which is given by Eq.~(\ref{eq2-2-7}).
The expression~(\ref{eq4-6}) depends on $\beta\equiv\ln(t/\sqrt{\xi}\delta_s)$ which comes from $\mu_{\mathrm{str}}$ in Eq.~(\ref{eq2-2-6}).
Here we use the value $\beta\simeq 4$ obtained by using parameters which are used in numerical simulations.
Both of the time scales $\tau_1$ and $\tau_2$ become comparable, but $\tau_2$ is slightly shorter than $\tau_1$
because of the small value of $\beta$.
We summarize the typical time scales given by Eqs.~(\ref{eq4-2}), (\ref{eq4-3}) and (\ref{eq4-6}) in Table~\ref{tab2}.

{\tabcolsep = 5mm
\begin{table}[h]
\begin{center}
\caption{Typical time scales for various values of $\kappa$.}
\vspace{3mm}
\begin{tabular}{c c c c}
\hline\hline
$\kappa$ & $\tau_2$ & $\tau_1$ & $\tau_a$ \\ 
\hline
0.4 & 3.15 & 3.20 & 3.29 \\
0.35 & 3.59 & 3.65 & 3.76 \\
0.3 & 4.19 & 4.27 & 4.38 \\
0.25 & 5.03 & 5.12 & 5.26 \\
0.2 & 6.29 & 6.40 & 6.57 \\
\hline\hline
\label{tab2}
\end{tabular}
\end{center}
\end{table}
}

\begin{figure}[htbp]
\begin{center}
\begin{tabular}{c c}
\resizebox{70mm}{!}{\includegraphics[angle=0]{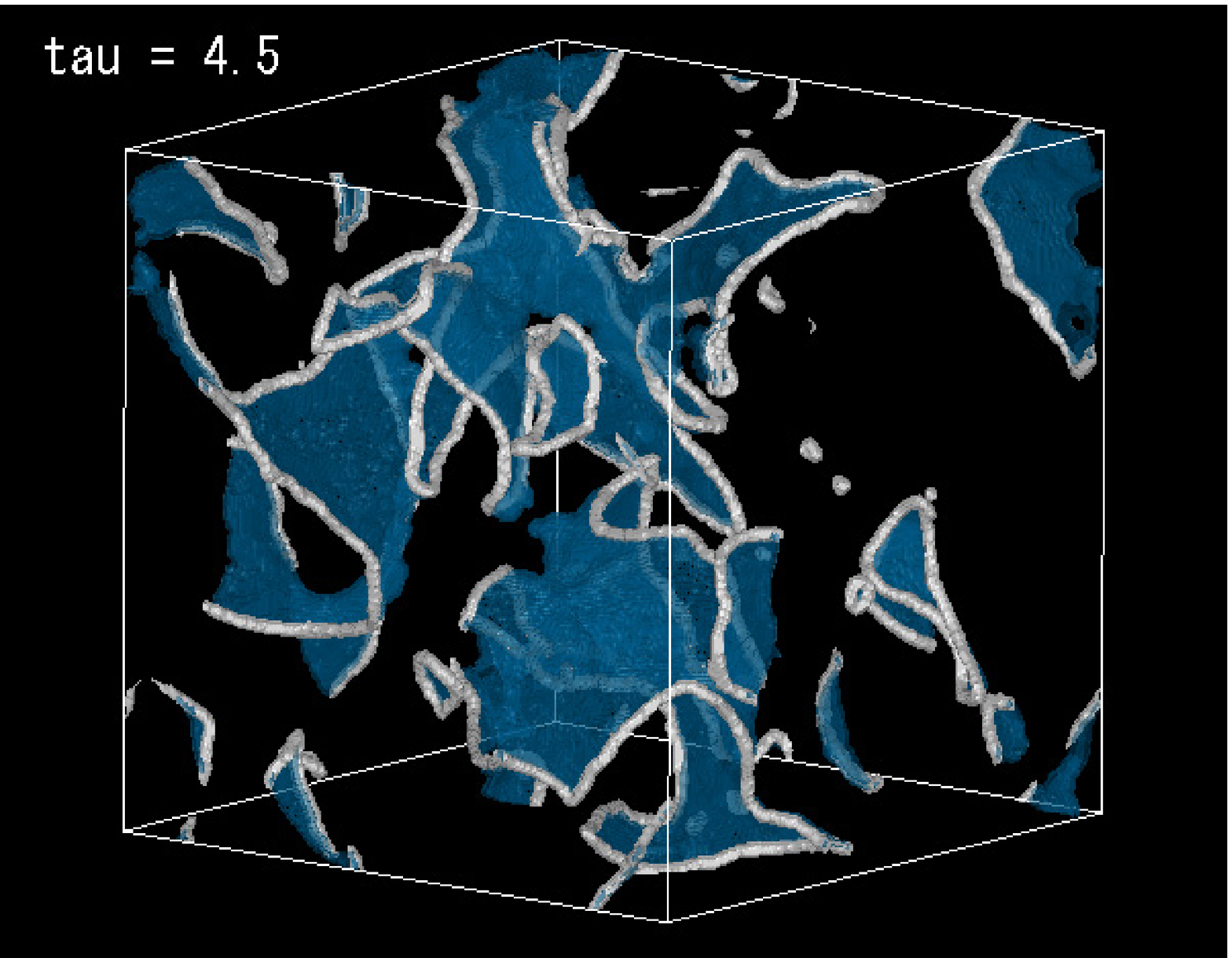}} &
\resizebox{70mm}{!}{\includegraphics[angle=0]{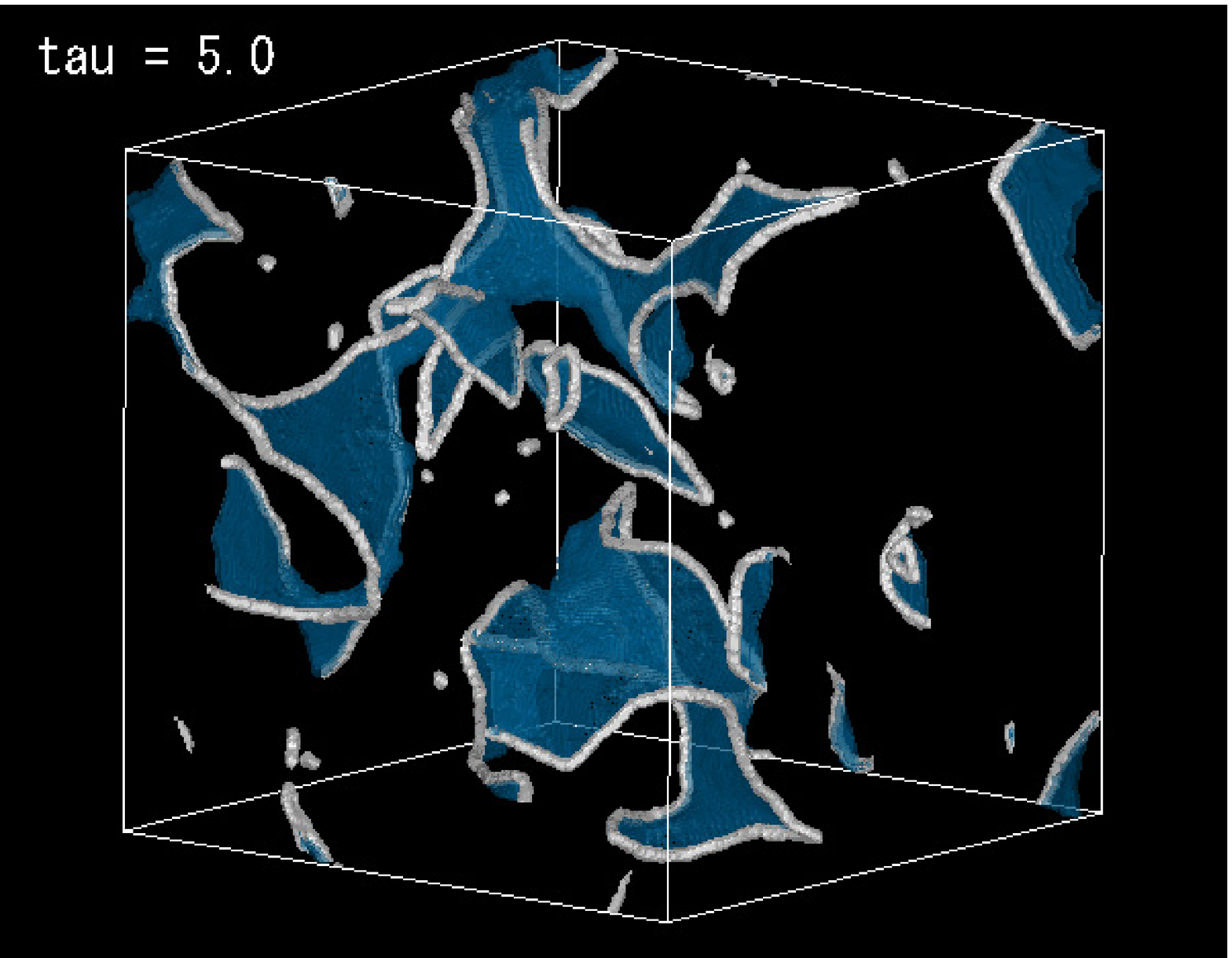}} \\
\resizebox{70mm}{!}{\includegraphics[angle=0]{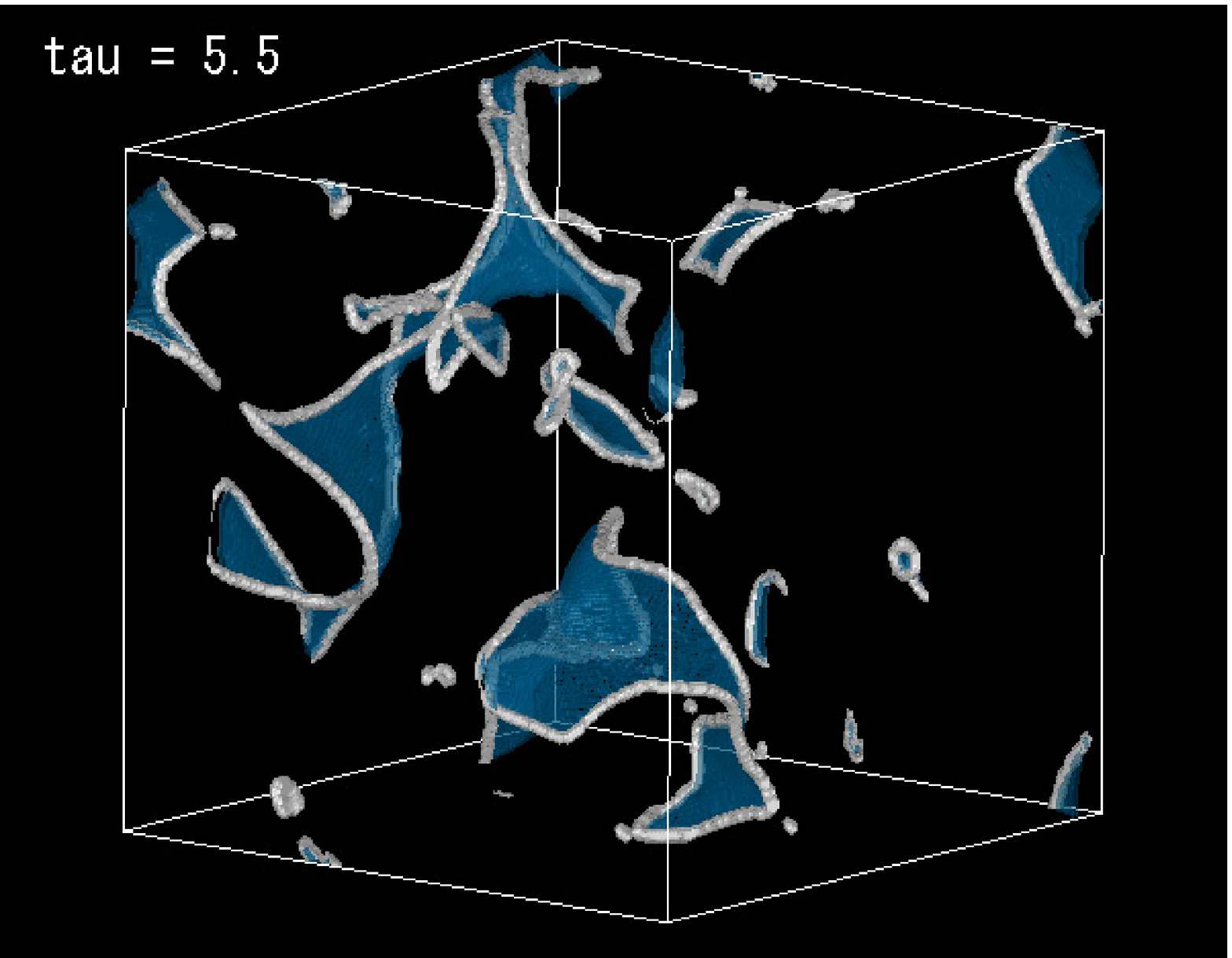}} &
\resizebox{70mm}{!}{\includegraphics[angle=0]{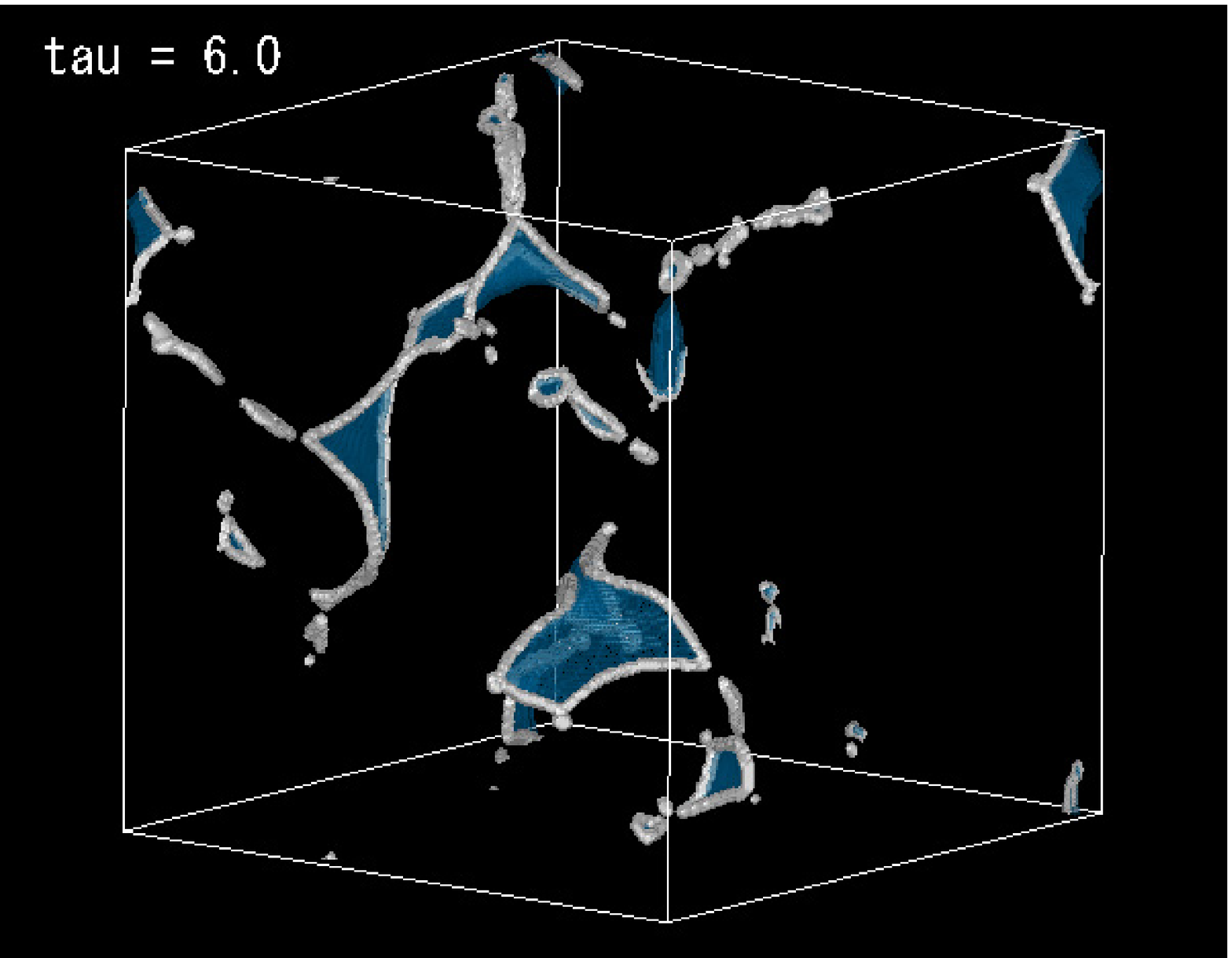}} \\
\resizebox{70mm}{!}{\includegraphics[angle=0]{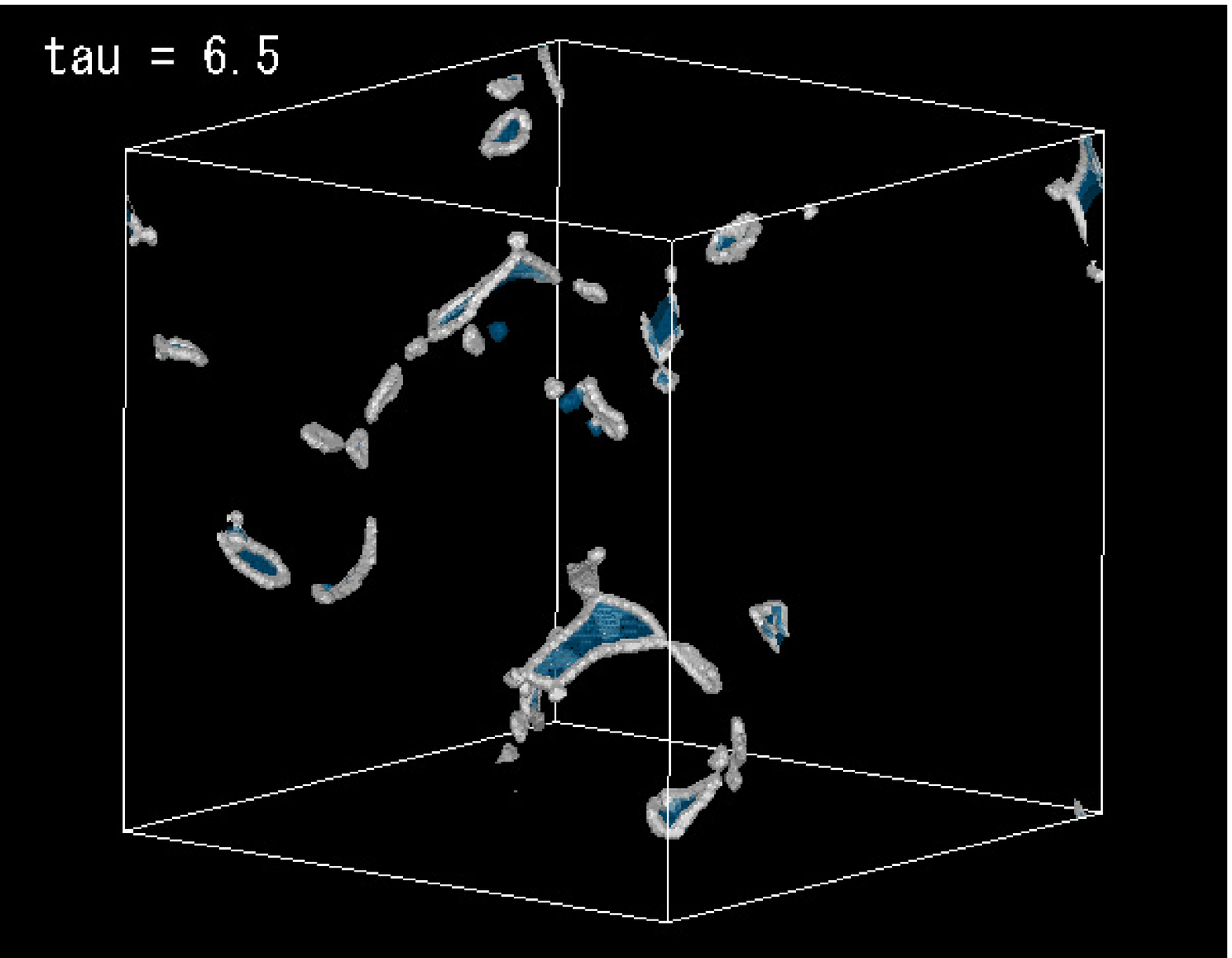}} &
\resizebox{70mm}{!}{\includegraphics[angle=0]{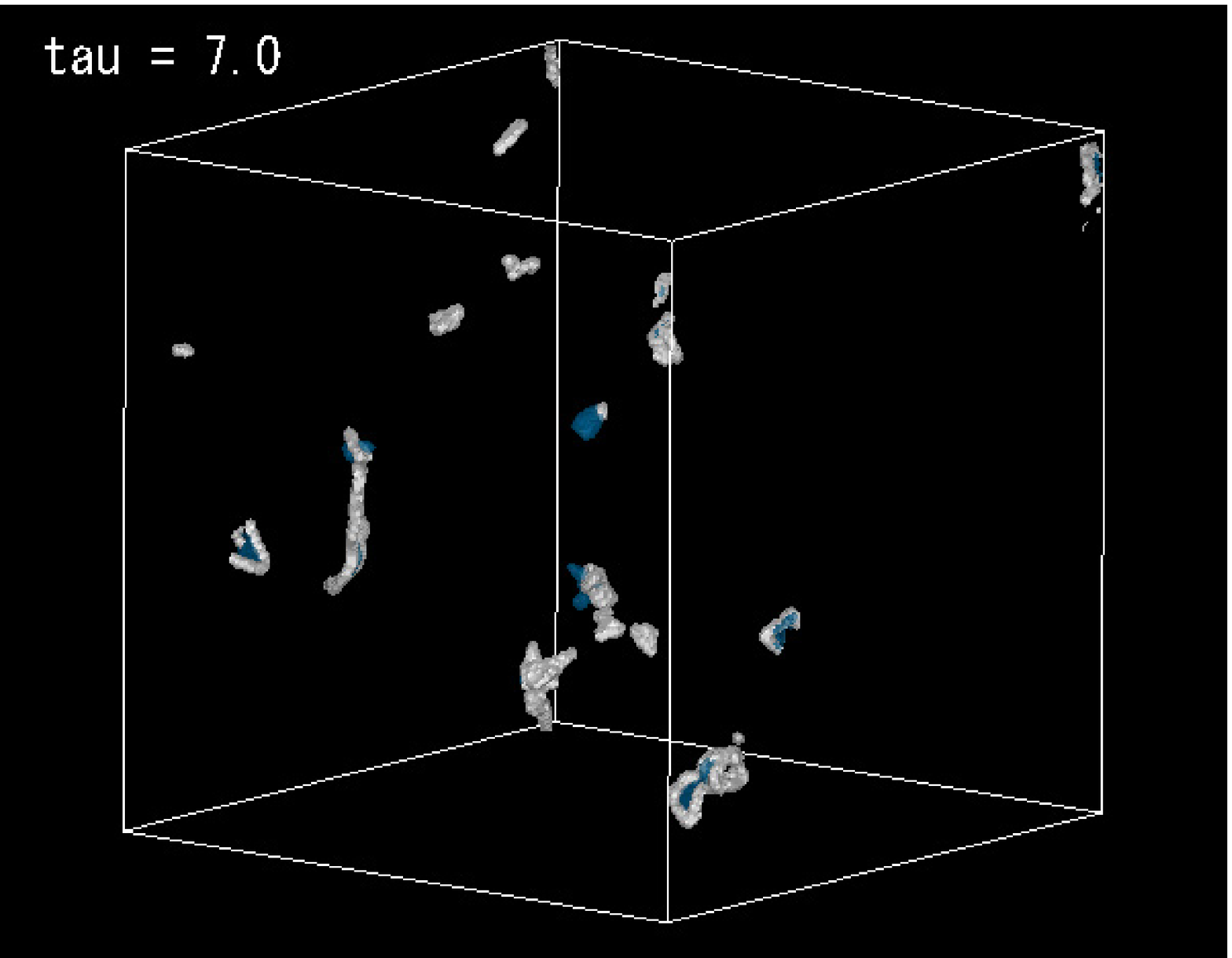}} \\
\end{tabular}
\end{center}
\caption{Visualization of one realization of the simulation.
In this figure, we take the box size as $L=15$ and $N=256$, which is smaller than that shown in Table~\ref{tab1}.
Other parameters are fixed so that $\lambda=1.0$, $\zeta=3.0$, $\alpha=2.0$, and $\kappa=0.4$.
The white lines correspond to the position of strings, while the blue surfaces correspond to the position of the center of domain walls.}
\label{fig2}
\end{figure}

\begin{figure}[htbp]
\begin{center}
\includegraphics[scale=0.5]{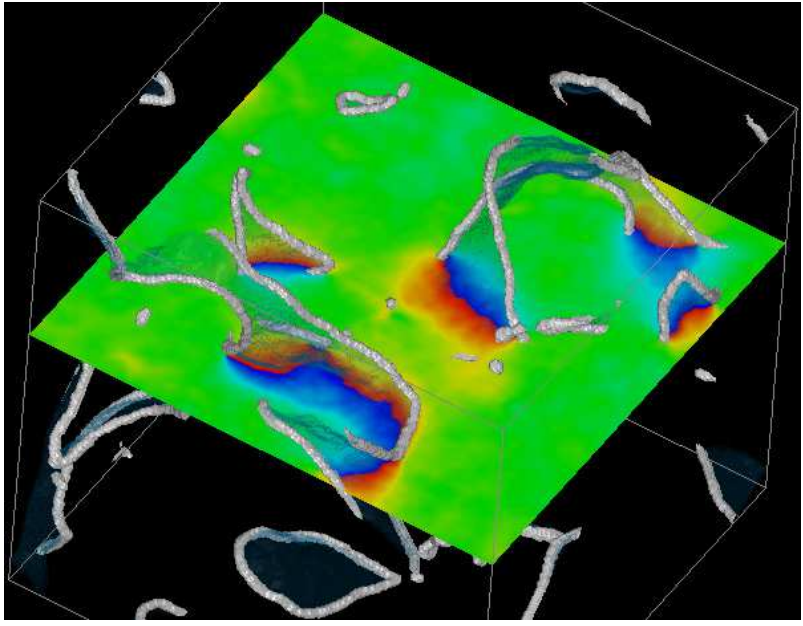}
\end{center}
\caption{The distribution of the phase of the scalar field $\theta$ on the two-dimensional slice of the simulation box.
In this figure, we used the same data that are used to visualize the result with $\tau = 5.0$ in Fig.~\ref{fig2}.
The value of $\theta$ varies from $-\pi$ (blue) to $\pi$ (red).
Domain walls are located around the region on which $\theta$ passes through the value $\pm\pi$,
while the green region corresponds to the true vacuum ($\theta=0$).
The length scale of the change of $\theta$ is roughly estimated as $\sim m_a^{-1}$,
which gives the thickness of domain walls.}
\label{fig3}
\end{figure}

Now, let us show the results of the simulations.
Figure~\ref{fig2} shows the visualization of one realization of the simulation. 
We see that at the first stage of the simulation, strings evolve and keep the scaling property.
However, at late time they shrink because of the tension of domain walls.
We also show the spatial distribution of the phase of the scalar field $\theta$ in Fig.~\ref{fig3}.
Note that the width of domain walls $\sim m_a^{-1}$ is much greater than that of strings $\sim (\sqrt{\lambda}\eta)^{-1}$, as shown in Fig.~\ref{fig3}.

\begin{figure}[htbp]
\begin{center}
\begin{tabular}{c c}
\resizebox{85mm}{!}{\includegraphics[angle=0]{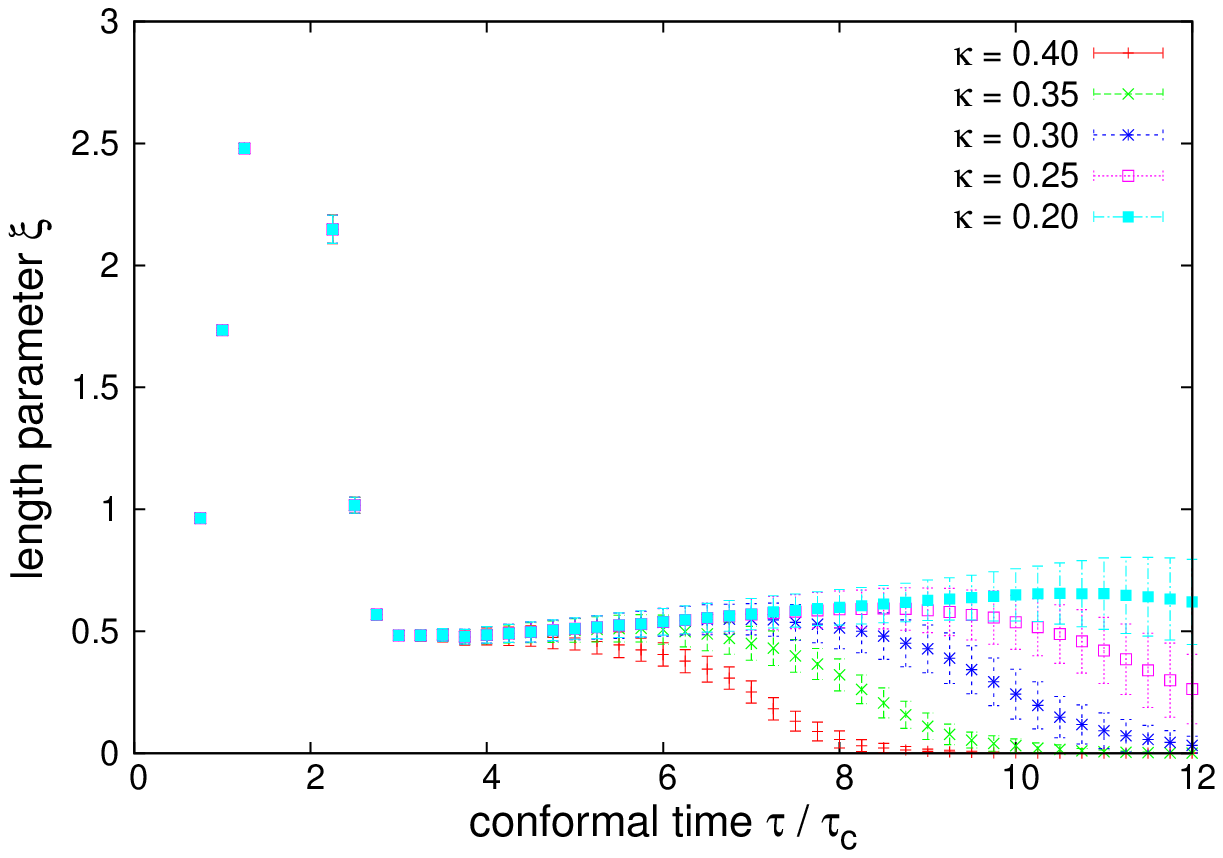}} &
\resizebox{85mm}{!}{\includegraphics[angle=0]{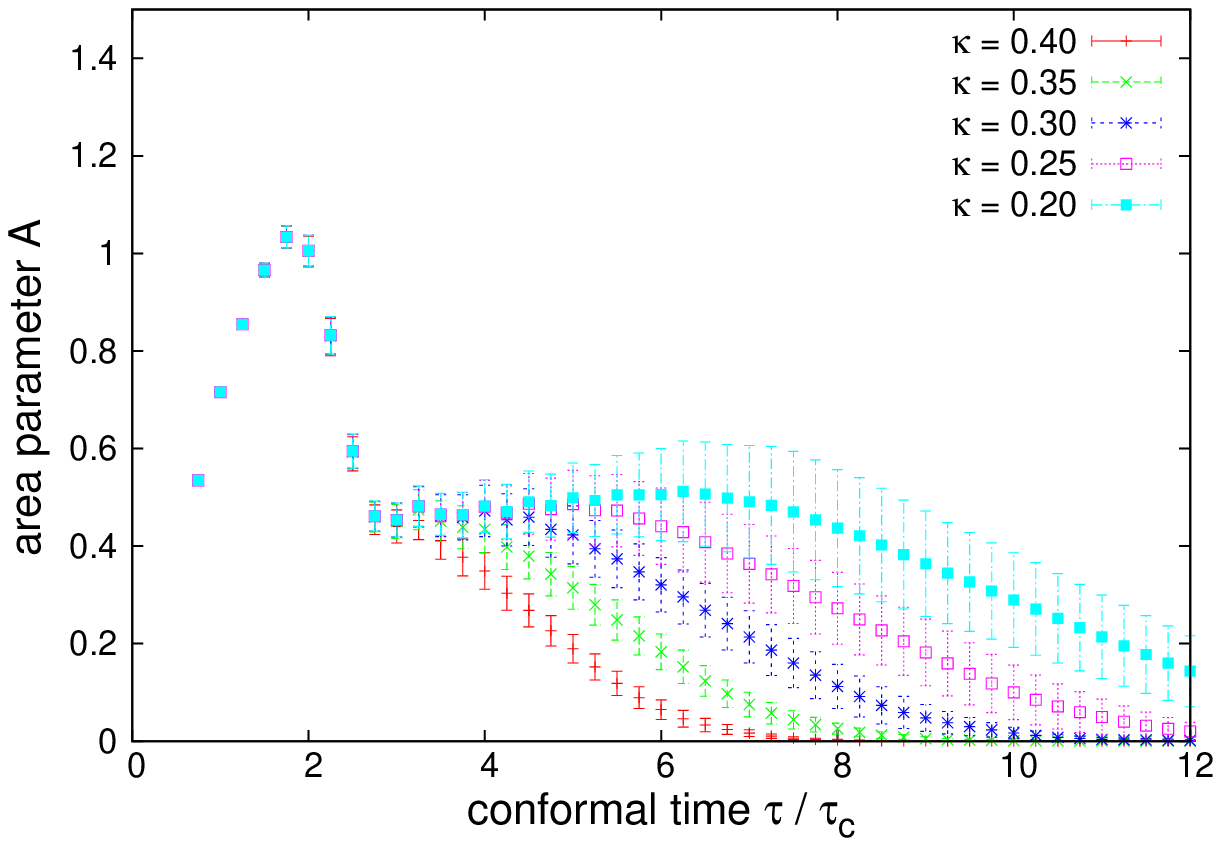}} \\
\end{tabular}
\end{center}
\caption{Time evolution of the length parameter $\xi$~(left panel) and the area parameter ${\cal A}$~(right-hand panel) for various values of $\kappa$.
Although walls do not exist before the time $\tau_1$, we can show the value of ${\cal A}$ evaluated at the time $\tau<\tau_1$.
This is because the value of ${\cal A}$ is calculated from the number of grid points on which the phase of the scalar filed passes the value $\theta=\pi$.
In this sense, ${\cal A}$ represents the area of domain walls only after the time $\tau_1$.}
\label{fig4}
\end{figure}

We performed 20 realizations for each choice of the parameter $\kappa$.
For each realization, we calculated the length parameter of strings
\begin{equation}
\xi \equiv \frac{\rho_{\mathrm{string}}}{\mu_{\mathrm{string}}}t^2, \label{eq4-7}
\end{equation}
and the area parameter of domain walls
\begin{equation}
{\cal A} \equiv \frac{\rho_{\mathrm{wall}}}{\sigma_{\mathrm{wall}}}t. \label{eq4-8}
\end{equation}
Figure~\ref{fig4} shows the time evolutions of $\xi$ and ${\cal A}$ for various values of $\kappa$.
Comparing the plot of ${\cal A}$ with Table~\ref{tab2}, we see that the value of ${\cal A}$
deviates from the scaling behavior (${\cal A}\simeq$constant) and begins to fall off around $\tau=\tau_1\simeq\tau_2$.
Note that $\xi$ starts to fall later than ${\cal A}$ does. This can be interpreted as follows.
Since domain walls are two-dimensional objects, they curve in various directions.
This curvature gets stretched when the tension of walls becomes effective.
The stretching process of walls reduces the value of ${\cal A}$, but might not affect the length of strings (i.e. the value of $\xi$).
Later, stretched walls pull the strings attached on their boundaries, which causes the reduction of $\xi$.

\begin{figure}[htbp]
\begin{center}
\includegraphics[scale=1.0]{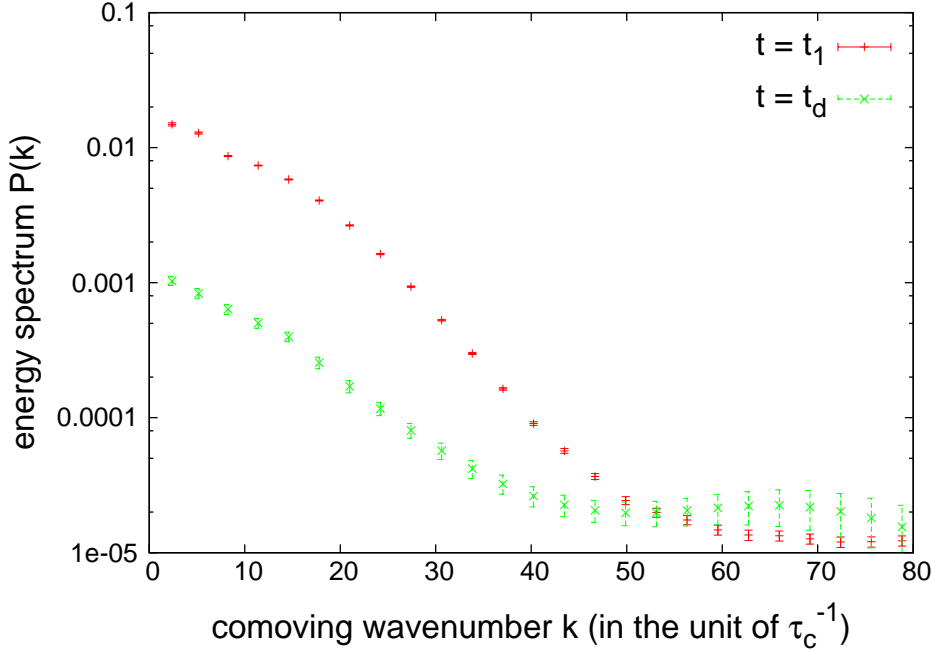}
\end{center}
\caption{The power spectrum of free axions calculated by using PPSE formula~(\ref{eq3-3-8}) in the simulations with $\kappa=0.3$.
We plot the spectra evaluated at two different times $t_1$ and $t_d$.
Note that the result of $P(k,t_1)$ shown here does not contain the numerical factor defined in Eq.~(\ref{eq3-4-4})}
\label{fig5}
\end{figure}

\begin{figure}[htbp]
\begin{center}
\includegraphics[scale=1.0]{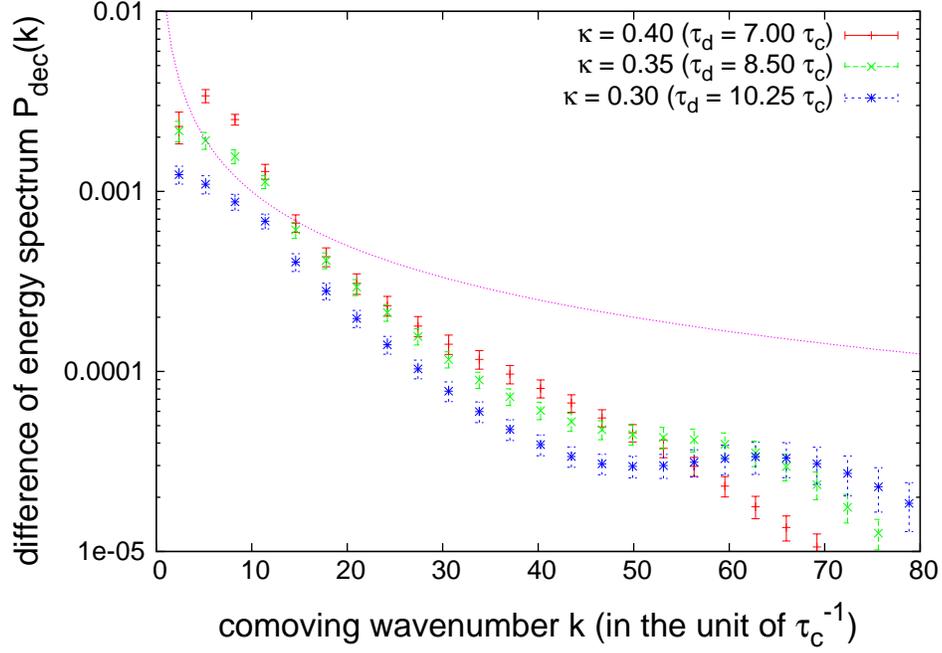}
\end{center}
\caption{The spectrum of axions produced by the decay of networks, defined by Eq.~(\ref{eq3-4-5}).
Note that the results with different value of $\kappa$ are evaluated at different times ($\tau_d$).
The form of the spectra is different from the relation $P_{\mathrm{dec}}(k)\propto 1/k$ which is indicated by the dotted line.}
\label{fig6}
\end{figure}

We also calculated the spectrum of axions radiated from strings and domain walls, using the method described in the previous section.
Figure~\ref{fig5} shows the spectra of free axions evaluated at $t_1$ and $t_d$.
The basic behavior of the spectrum evaluated at $t_1$ is similar to that obtained in Ref.~\cite{2011PhRvD..83l3531H}.
This spectrum is dominated by the contribution of axions produced by strings.
However, the population of axions with high momenta increases after the decay of domain walls ($t=t_d$).
The final form of the spectrum, obtained by subtracting the components of radiations produced before $t_1$, is shown in Fig.~\ref{fig6}.
The spectrum has a peak at the low momentum. This disagrees with the result of Chang, Hagmann, and Sikivie~\cite{1999PhRvD..59b3505C},
which claims that the radiated axions have a spectrum proportional to $1/k$.
Note that, however, there is a high frequency tail in the spectrum, which has a cutoff at the momentum corresponding to (twice the size of) the width of strings
$k\simeq (2\pi/2\delta_s)R(t_d) \simeq 64.4$ (for $\kappa=0.3$).
This feature might be interpreted in terms of the reasoning of~\cite{1987PhLB..195..361H,1991NuPhB.363..247H,1999PhRvD..59b3505C} (scenario B).
Namely, there are various size of defects around the time $t=t_d$, and small-scale defects can radiate axions with harder momenta.
As we see in Fig.~\ref{fig6}, the contribution from these hard axions is subdominant, and most axions have a momentum comparable to the mass of the axion $k/R(t_d)\sim m_a$.

Using the result of $P_{\mathrm{dec}}(k,t_d)$, we compute the mean comoving momentum of radiated axions
\begin{equation}
\bar{k}(t_d) = \frac{\int\frac{dk}{2\pi}P_{\mathrm{dec}}(k,t_d)}{\int\frac{dk}{2\pi}\frac{1}{k}P_{\mathrm{dec}}(k,t_d)}. \label{eq4-9}
\end{equation}
We define the ratio of the physical momentum $\bar{k}/R(t_d)$ to the axion mass $m_a(t_d)$, 
\begin{equation}
\epsilon_w \equiv \frac{\bar{k}(t_d)/R(t_d)}{m_a(t_d)}. \label{eq4-10}
\end{equation}
From the result of the numerical simulations with $\kappa=0.3$, we obtain
\begin{equation}
\bar{k}(t_d) = 5.76\pm0.15 \quad\mathrm{and}\quad \epsilon_w = 3.12\pm0.08. \label{eq4-11}
\end{equation}
This result corresponds to the mean energy of axions [cf. Eq.~(\ref{eq3-4-2})],
\begin{equation}
\bar{\omega}_a(t_d)/m_a(t_d) = \sqrt{1+\epsilon_w^2} = 3.28\pm0.08. \label{eq4-12}
\end{equation}

\begin{figure}[htbp]
\begin{center}
\includegraphics[scale=1.0]{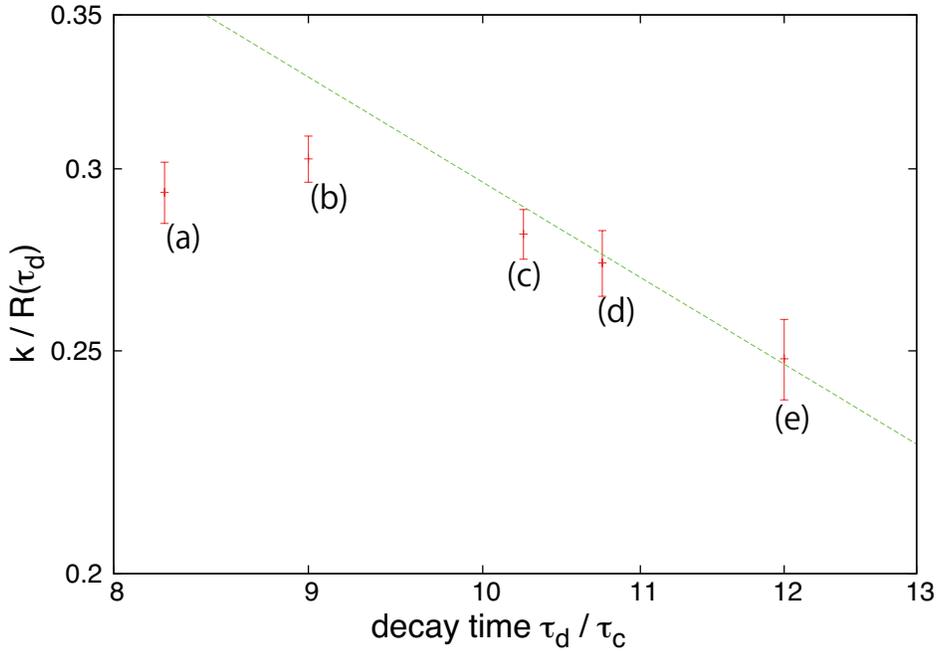}
\end{center}
\caption{The value of the mean physical momentum of radiated axions $\bar{k}/R(\tau_d)$ and its dependence on the choice of $\tau_d$.
The dashed line represents the relation $\bar{k}/R(\tau_d) \propto 1/R(\tau_d) \propto 1/\tau_d$. The five points correspond to the results with the
value of the area parameters (a) ${\cal A}(\tau_d)\lesssim$0.1, (b) ${\cal A}(\tau_d)\lesssim$0.05, (c) ${\cal A}(\tau_d)\lesssim$0.01, (d) ${\cal A}(\tau_d)\lesssim$0.005,
and (e) ${\cal A}(\tau_d)\lesssim$0.001.}
\label{fig7}
\end{figure}

Note that there are some ambiguities in this analysis.
For instance, we choose the time $t_d$, at which the decay of networks completes, by hand.
If we choose $t_d$ as sufficiently late time (for example, the final time of the simulations),
we would underestimate the mean momentum of radiated axions defined by Eqs.~(\ref{eq4-9}) and (\ref{eq4-10}),
since the physical momentum gets redshifted proportionally to $1/R(\tau_d)$.
Figure~\ref{fig7} shows the results of the physical momentum $\bar{k}/R(\tau_d)$ for various choices of $\tau_d$ in the simulations with $\kappa=0.3$.
The value of ${\bar k}/R(\tau_d)$ begins to shift as $\propto 1/R$ at the result with $\tau_d=10.25$.
This value of $\tau_d$ corresponds to the time at which the area of domain walls becomes less than ${\cal O}(1)\%$ of the Hubble scale (${\cal A}\lesssim$0.01).
Therefore, we choose $\tau_d$ as the time at which the value of ${\cal A}$ falls below 0.01.
We use this criterion to calculate the spectra with different values of $\kappa$ shown in Fig.~\ref{fig6}.

Another subtlety is whether the results of numerical simulations are sensitive to the choice of $\kappa\equiv \Lambda/F_a$.
There is a tremendous hierarchy between the QCD scale and the PQ scale, $\Lambda/F_a\simeq 100\mathrm{MeV}/10^{10}\mathrm{GeV}=10^{-11}$,
but we cannot perform the simulations with such a small value of $\kappa$ because of the limitation of the dynamical range.
Nonetheless, we believe that the ratio between the mean momentum of the radiated axions and the typical momentum scale (such as $m_a$)
is not so sensitive to the value of $\kappa$, since the power spectrum has a sharp peak at the typical momentum scale as we see in Fig.~\ref{fig6}.
However, one might regard the peak momentum as the inverse of the horizon scale $\sim 2\pi/t_d$, instead of the mass of the axion.
The wrong interpretation of the peak momentum affects the estimation of the present abundance of axions, which will be given in the next section.
We present the result of the ratios $[\bar{k}/R(t_d)]/m_a(t_d)$ and $[\bar{k}/R(t_d)]/(2\pi/t_d)$ for $\kappa=0.3$, 0.35, and 0.4 in Table~\ref{tab3}.
We observe that the $\kappa$ dependence of $[\bar{k}/R(t_d)]/m_a(t_d)$ is weaker than that of $[\bar{k}/R(t_d)]/(2\pi/t_d)$.
This fact implies that it is reasonable to assume $\bar{k}/R(t_d)\propto m_a(t_d)$ rather than $\bar{k}/R(t_d)\propto 2\pi/t_d$.
For now, it is difficult to discuss the significance of the result in which the value of $[\bar{k}/R(t_d)]/m_a(t_d)$ varies with $\kappa$, because of the lack of samples.
It can be said that the result with larger values of $\kappa$ is unreliable since the networks of strings begin to collapse soon after their formation.
Indeed, in Fig.~\ref{fig6} we observe that the high-momentum cutoff, which is the consequence of the ``smallness'' of $\kappa$ or $m_a$ compared with $F_a$,
is less apparent in the result with $\kappa=0.4$. This ambiguity should be resolved in future numerical studies with improved dynamical ranges.
Regarding this ambiguity, we use the result with the smallest value of $\kappa$ given by Eqs.~(\ref{eq4-11}) or (\ref{eq4-12}).

\begin{table}[h]
\begin{center}
\caption{Ratio between the mean momentum of radiated axions $\bar{k}/R(t_d)$ and the mass of the axion $m_a(t_d)$,
or the mean momentum of radiated axions $\bar{k}/R(t_d)$ and the inverse of the horizon scale $2\pi/t_d$, for different values of $\kappa$.}
\vspace{3mm}
\begin{tabular}{c c c}
$\kappa$ & $[\bar{k}/R(t_d)]/m_a(t_d)$ & $[\bar{k}/R(t_d)]/(2\pi/t_d)$\\
\hline
0.3 & 3.12$\pm$0.08 & 4.70$\pm$0.12\\
0.35 & 2.64$\pm$0.09 & 3.71$\pm$0.13\\
0.4 & 2.51$\pm$0.10 & 3.13$\pm$0.13\\
\label{tab3}
\end{tabular}
\end{center}
\end{table}

\begin{figure}[htbp]
\centering
$\begin{array}{cc}
\subfigure[]{
\includegraphics[width=0.45\textwidth]{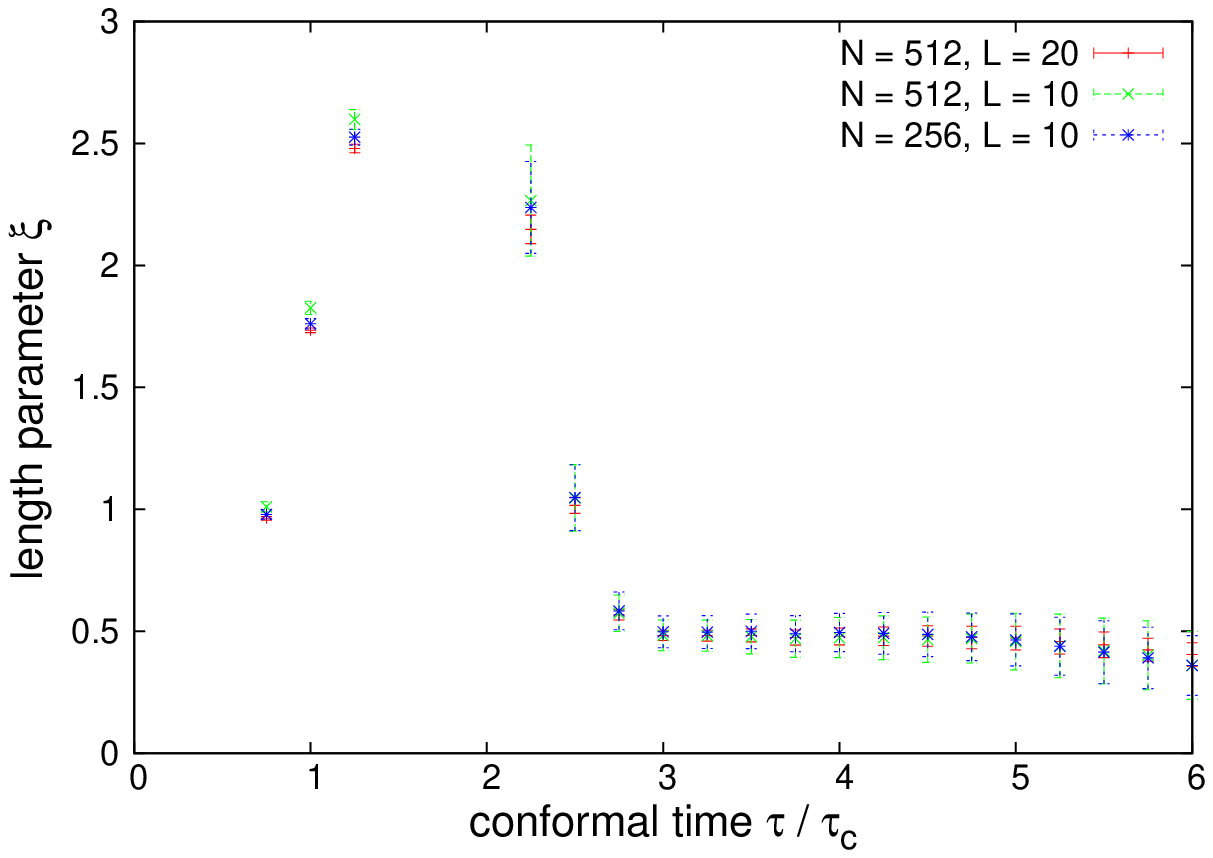}}
\hspace{20pt}
\subfigure[]{
\includegraphics[width=0.45\textwidth]{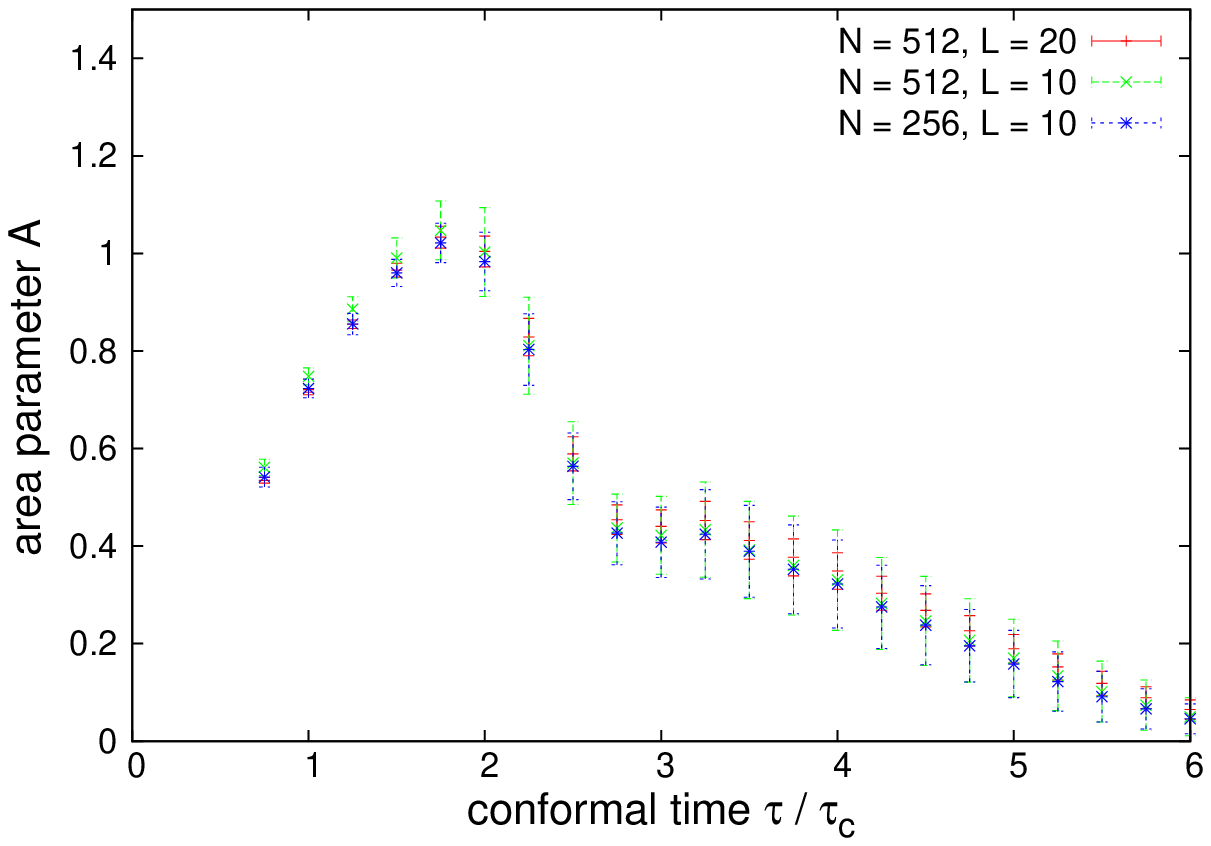}}
\end{array}$
\subfigure[]{
\includegraphics[width=0.45\textwidth]{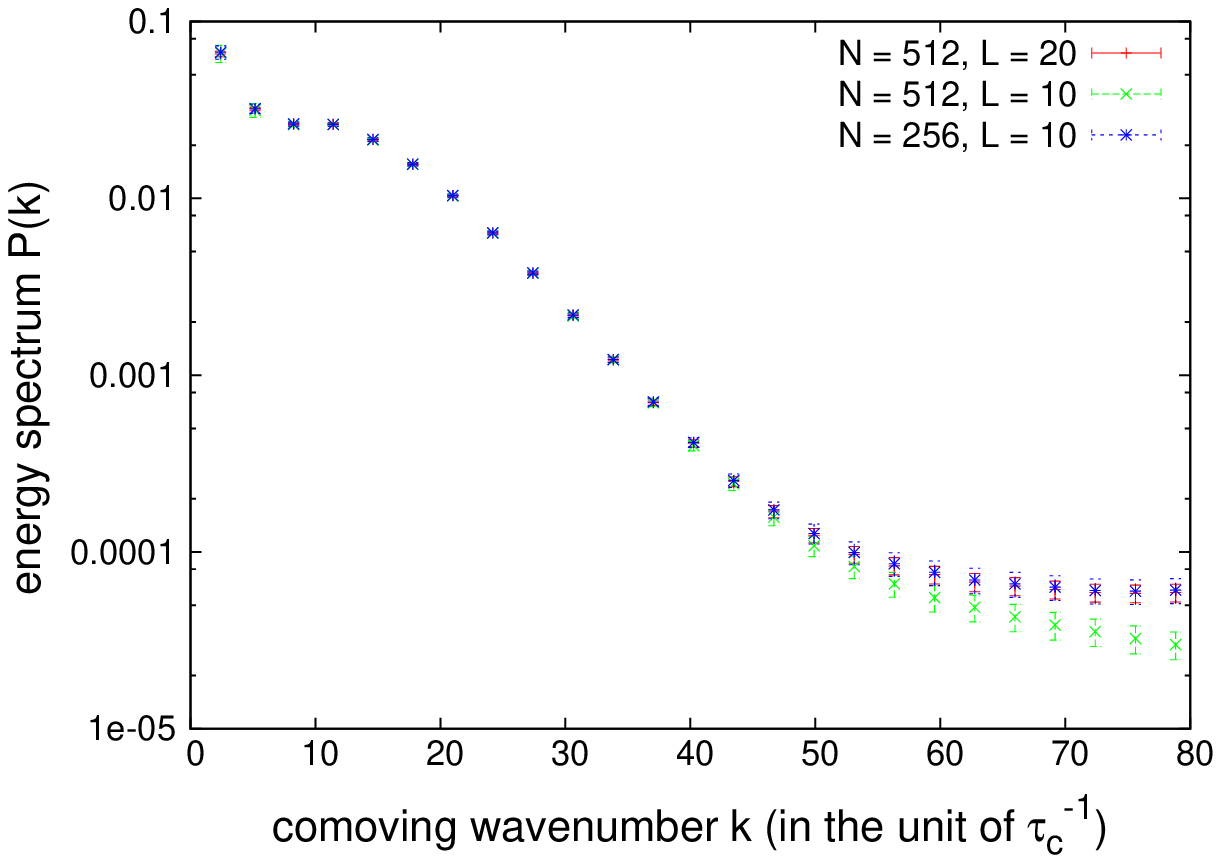}}
\caption{The comparison between simulations with larger dynamical range ($N=512$, $L=20$) and that with smaller dynamical range ($N=512$, $L=10$) and ($N=256$, $L=10$)
on the result of (a) time evolution of the length parameter, (b) time evolution of the area parameter and (c) power spectrum of free axions evaluated at time $t_1$.
In these simulations, we choose the same set of parameters shown in Table~\ref{tab1} except that $\kappa=0.4$.
Note that we cannot calculate the difference of power spectrum $P_{\mathrm{dec}}(k,t_d)$ and the mean momentum $\bar{k}(t_d)$,
since the time $t_d$ is beyond the final time in the simulation with smaller dynamical range.}
\label{fig8}
\end{figure}

The limitation of the dynamical range of the simulation is provided by the conditions described above Eq.~(\ref{eq4-1}). Especially, the condition that
the width of strings $\delta_s$ should be grater than the lattice spacing $\delta x_{\mathrm{phys}}$ might be marginally violated at the end of the simulation.
We are not confident that our choice, $\delta_s/\delta x_{\mathrm{phys}} \simeq 1.07$ at the end of the simulation, is safe enough.
To clarify this point, we performed the set of test simulations with smaller dynamical range and larger string width by tuning the grid size $N$ and the box size $L$.
We chose two set of parameters, ($N=512$, $L=10$) and ($N=256$, $L=10$), corresponding to the string width
$\delta_s/\delta x_{\mathrm{phys}}|_{\tau=6}\simeq 4.27$ and $2.13$, respectively.
Since the box size $L=10$ is half of the full simulation with $L=20$, we cannot run the simulation beyond the time $\tau=6$, otherwise
the condition on the Hubble radius described above Eq.~(\ref{eq4-1}) might be violated.
In Fig.~\ref{fig8}, we compare the results of these test simulations with the results found in simulations with larger dynamical range ($N=512$, $L=20$).
We confirmed that there is no dramatic change in results of the time evolution of scaling parameters, except that the error bars become slightly larger for the simulation with a smaller dynamical range.
This result quantitatively supports our supposition that the dynamical range does not much affect the results of the simulations.
However, as we see in Fig.~\ref{fig8} (c), the result of the energy spectrum in the simulation with high resolution ($N=512$, $L=10$) deviates from others at wavenumber $k\gtrsim 50$.
This indicates that at wavenumber greater than $k\sim 50$ the result depends on the spatial resolution, but the result is robust at small wavenumber $k\lesssim 50$.
We may overestimate the abundance of small scale modes, but it does not much affect the final results since their contribution is subdominant.

Furthermore, there are ambiguities in the values of scaling parameters defined in Eqs.~(\ref{eq4-7}) and (\ref{eq4-8}).
Our result $\xi\simeq 0.5$, shown in Fig.~\ref{fig4}, is somewhat lower than the value $\xi\simeq 0.8$-1.3
obtained in previous studies~\cite{1999PhRvL..82.4578Y,1999PhRvD..60j3511Y,2003PhRvD..67j3514Y,2011PhRvD..83l3531H}.
This might be caused by the different choice of the parameter $\zeta$ used as an input of the numerical simulations.
Our choice $\zeta=3.0$ is smaller than the values $\zeta= 8$-10 used in past numerical simulations~\cite{1999PhRvL..82.4578Y,1999PhRvD..60j3511Y,2003PhRvD..67j3514Y}.
The parameter $\zeta$ controls the magnitude of the symmetry breaking scale $\eta$ [see Eq.~(\ref{eq3-1-16})], which determines the width of global stings $\delta_s\propto 1/\eta$.
Therefore, different choice of $\zeta$ affects the emission rate of Nambu-Goldstone bosons from strings~\cite{1994csot.book.....V,2002PhRvD..65d3514M}
$\Gamma_{\mathrm{NG}} = \tilde{\Gamma}/[2\pi L_s\ln(L_s/\delta_s)]$, where $\tilde{\Gamma}$ is a numerical factor of ${\cal O}$(10-100) and $L_s\sim t$ is the characteristic length scale of strings.
The simulation with small value of $\zeta$ corresponds to the simulation with thick strings, in which the global string networks lose their energy efficiently due to
the emission of Nambu-Goldstone bosons, since the logarithmic correction to the emission rate $\Gamma_{\mathrm{NG}}\propto1/\ln(t/\delta_s)$ becomes large.
This large emission rate of Nambu-Goldstone bosons reduces the energy density of global string networks and suppresses the value of $\xi$.
However, it was argued that in the realistic case with $\ln(t/\delta_s)\approx 70$ the radiative effect becomes subdominant, and the value of $\xi$
is purely determined by the formation rate of loops~\cite{2002PhRvD..65b3503M,2003PhRvD..67j3514Y}.
Regarding this effect, the authors of Ref.~\cite{2003PhRvD..67j3514Y} estimated the final value of scaling parameter as $\xi=1.6\pm0.3$.
Indeed, the results with smaller values of $\kappa$ in fig.~\ref{fig4} indicate that the value of $\xi$ increases due to the change of emission rate $\Gamma_{\mathrm{NG}}\propto1/\ln(t/\delta_s)$
with time. We anticipate that the value of the length parameter gradually reaches the final value $\xi\approx 1$, which cannot be observed in the simulations with the limited dynamical range.

The reason why we choose a smaller value of $\zeta$ is to improve the dynamical range of simulations by keeping the width of strings greater than the lattice spacing [see Eq.~(\ref{eq4-1})].
This choice enables us to try to perform simulations with varying the values of $\kappa$ but invalidates the estimation of $\xi$ due to the large emission rate of Nambu-Goldstone bosons.
However, we believe that this choice does not affect our main result about the radiated axions produced by domain walls, since the choice of the parameter $\zeta$ only controls the
small-scale properties such as the width of strings, while the population of axions is dominated by low-momentum modes which are governed by the large-scale physics with the wavelength
comparable to the inverse of the axion mass.

The precise determination of the values of scaling parameters including the effect of backreaction of Numbu-Goldstone boson emissions is beyond the scope of this paper.
To be conservative, we use the rough estimate $\xi\simeq 1.0\pm 0.5$ with $50\%$ uncertainty when we calculate the abundance of cold axions in the next section.
We also assume that the area parameter ${\cal A}$ possesses similar uncertainty, and use the value ${\cal A}\simeq 0.50\pm 0.25$ around the time of the formation of domain walls.
\section{\label{sec5}Relic abundance of cold dark matter axions}
In this section, we calculate the abundance of cold dark matter axions. These axions are produced in
three processes: (a) vacuum misalignment, (b) string decay and (c) domain wall decay. In the following,
we treat these three cases, respectively, and estimate the energy density of axions at the present time.
\subsection{\label{sec5-1}Zero modes}
The averaged homogeneous value of the axion field (zero mode) begins to oscillate around the minimum of the potential at the time $t_1$.
Let us denote the initial amplitude of the $\theta$ angle as $\theta_1$.
The energy density of the zero modes is given by
\begin{equation}
\rho_{a,0}(t_1) = \frac{1}{2}m_a(T_1)^2\theta_1^2F_a^2. \label{eq5-1-1}
\end{equation}
Noting that the number of zero mode axions is fixed after the time $t_1$, we find the energy density at the present time $t_0$
\begin{equation}
\rho_{a,0}(t_0) = \rho_{a,0}(t_1)\frac{m_a(0)}{m_a(T_1)}\left(\frac{R(t_1)}{R(t_0)}\right)^3. \label{eq5-1-2}
\end{equation}
From the entropy conservation, it follows that
\begin{equation}
\left(\frac{R(t_1)}{R(t_0)}\right)^3 = \frac{s_0}{\frac{2\pi^2}{45}g_{*,1}T_1^3}, \label{eq5-1-3}
\end{equation}
where $s_0$ is the entropy density at the present time, which satisfies
\begin{equation}
\frac{s_0h^2}{\rho_{c,0}} = \frac{4}{3}\frac{g_{*S,0}}{g_{*,0}}\frac{\Omega_Rh^2}{T_0}. \label{eq5-1-4}
\end{equation}
Here, $\rho_{c,0}$ is the critical density today, $g_{*S,0}$ and $g_{*,0}$ are the effective degrees of freedom for entropy density and
energy density of radiations at the present time~\cite{1990eaun.book.....K}, $T_0$ is the temperature today, $\Omega_Rh^2 \equiv \rho_R(t_0)h^2/\rho_{c,0}$ is
the density parameter of radiations, and $h$ is the reduced Hubble parameter: $H_0=100h$km$\cdot$sec$^{-1}$Mpc$^{-1}$.
Using Eqs.~(\ref{eq5-1-1}) - (\ref{eq5-1-4}) and the expression for $T_1$ given by Eq.~(\ref{eq2-2-2}),
we find that the density parameter of the zero mode axion $\Omega_{a,0}h^2=\rho_{a,0}(t_0)h^2/\rho_{c,0}$ becomes
\begin{equation}
\Omega_{a,0}h^2 = 0.095\times \theta_1^2\left(\frac{g_{*,1}}{70}\right)^{-(n+2)/2(n+4)}\left(\frac{F_a}{10^{12}\mathrm{GeV}}\right)^{(n+6)/(n+4)}\left(\frac{\Lambda}{400\mathrm{MeV}}\right). \label{eq5-1-5}
\end{equation}

If PQ symmetry is broken after inflation, the value of $\theta_1$ spatially varies, and we can replace $\theta_1$ by the root-mean-square value
\begin{equation}
\langle\theta_1^2\rangle = \frac{1}{2\pi}\int^{\pi}_{-\pi}\theta_1^2d\theta_1 = \pi^2/3. \nonumber
\end{equation}
Furthermore, it was pointed out that the anharmonic effect becomes important for a large value of $\theta_1$~\cite{1986PhRvD..33..889T,1992PhRvD..45.3394L,2008JCAP...09..005B}.
Considering this effect, we take the replacement $\theta_1^2\to c_{\mathrm{anh}}\frac{\pi^2}{3}$, where $c_{\mathrm{anh}}$ is the anharmonic correction.
Turner~\cite{1986PhRvD..33..889T} calculated the anharmonic effect numerically and obtained the correction factor $c_{\mathrm{anh}}=1.9$-$2.4$.
Later, Lyth~\cite{1992PhRvD..45.3394L} gave the extensive calculation and reported the agreement with Turner's result within a factor of 2.
The more precise calculation given by~\cite{2008JCAP...09..005B} leads $c_{\mathrm{anh}}\simeq1.85$.
Here we take the value $\theta_1^2 \to 1.85\times \frac{\pi^2}{3}$, and obtain
\begin{equation}
\Omega_{a,0}h^2 = 0.58\times\left(\frac{g_{*,1}}{70}\right)^{-(n+2)/2(n+4)}\left(\frac{F_a}{10^{12}\mathrm{GeV}}\right)^{(n+6)/(n+4)}\left(\frac{\Lambda}{400\mathrm{MeV}}\right). \label{eq5-1-6}
\end{equation}
\subsection{\label{sec5-2}Axions radiated from scaling global strings}
We can estimate the energy density of axions radiated from strings by using a similar method introduced in Appendix B of~\cite{2011PhRvD..83l3531H}.
In~\cite{2011PhRvD..83l3531H}, it was assumed that the number of axions is fixed after the time $t_2$.
However, since the mass of the axion becomes non-negligible after the time $t_1$ (recall that $t_1<t_2$),
the analysis of~\cite{2011PhRvD..83l3531H} is applicable only for the radiations produced before the time $t_1$.
Here we assume that the radiation of axions from scaling global strings is terminated at $t_1$.
The abundance of axions radiated after $t_1$ will be estimated in the next subsection.

The present number density of axions radiated before $t_1$ is estimated as~\cite{2011PhRvD..83l3531H}
\begin{equation}
n_{a,\mathrm{str}}(t_0) = \left(\frac{R(t_1)}{R(t_0)}\right)^3\frac{F_a^2}{t_1}\frac{\xi}{\epsilon}\ln\left(\frac{t_1/\sqrt{\xi}}{\delta_s}\right), \label{eq5-2-1}
\end{equation}
where $\epsilon$ is the ratio between the mean momentum of axions radiated by strings and the horizon scale.

The energy density of axions today is given by $\rho_{a.\mathrm{str}}(t_0) = m_a(0)n_{a,\mathrm{str}}(t_0)$.
Then, we found the density parameter of axions radiated by strings
\begin{equation}
\Omega_{a,\mathrm{str}}h^2 = 8.74\times\frac{\xi}{\epsilon}\left(\frac{g_{*,1}}{70}\right)^{-(n+2)/2(n+4)}\left(\frac{F_a}{10^{12}\mathrm{GeV}}\right)^{(n+6)/(n+4)}\left(\frac{\Lambda}{400\mathrm{MeV}}\right). \label{eq5-2-2}
\end{equation}
In numerical simulations performed by Ref.~\cite{2011PhRvD..83l3531H}, the value of $\epsilon$ is estimated as $\epsilon^{-1}=0.23\pm0.02$.
By using this value for $\epsilon$ and the rough estimation for the length parameter $\xi\simeq1.0\pm 0.5$, we obtain
\begin{equation}
\Omega_{a,\mathrm{str}}h^2 = (2.0\pm1.0)\times\left(\frac{g_{*,1}}{70}\right)^{-(n+2)/2(n+4)}\left(\frac{F_a}{10^{12}\mathrm{GeV}}\right)^{(n+6)/(n+4)}\left(\frac{\Lambda}{400\mathrm{MeV}}\right). \label{eq5-2-3}
\end{equation}
\subsection{\label{sec5-3}Axions radiated from the decay of string-wall systems}
Define the area parameter of domain walls at $t=t_1$
\begin{equation}
{\cal A}_1 \equiv \frac{\rho_{\mathrm{wall}}(t_1)}{\sigma_{\mathrm{wall}}(t_1)}t_1, \label{eq5-3-1}
\end{equation}
and the length parameter of strings at $t=t_1$
\begin{equation}
\xi_1 \equiv \frac{\rho_{\mathrm{string}}(t_1)}{\mu_{\mathrm{string}}(t_1)}t_1^2. \label{eq5-3-2}
\end{equation}
The string-wall networks begin to collapse around the time $t=t_1$.
We simply assume that, after the time $t_1$, the whole energy stored in these defects is diluted as $R(t)^{-3}$ due to the cosmic expansion
\begin{equation}
\rho_{\mathrm{string}\mathchar`-\mathrm{wall}}(t) = \left[{\cal A}_1\frac{\sigma_{\mathrm{wall}}(t_1)}{t_1}+\xi_1\frac{\mu_{\mathrm{string}}(t_1)}{t_1^2}\right]\left(\frac{R(t_1)}{R(t)}\right)^3 \qquad \mathrm{for}\qquad t>t_1. \label{eq5-3-3}
\end{equation}
Suppose that the decay is complete at the time $t_d>t_1$.
The number density of axions produced by the decay of string-wall networks is
\begin{align}
n_{a,\mathrm{dec}}(t) &= \frac{\rho_{\mathrm{string}\mathchar`-\mathrm{wall}}(t_d)}{\bar{\omega}_a}\left(\frac{R(t_d)}{R(t)}\right)^3 \nonumber \\
&= \frac{1}{\sqrt{1+\epsilon_w^2}m_a(t_d)}\left[{\cal A}_1\frac{\sigma_{\mathrm{wall}}(t_1)}{t_1}+\xi_1\frac{\mu_{\mathrm{string}}(t_1)}{t_1^2}\right]\left(\frac{R(t_1)}{R(t)}\right)^3, \label{eq5-3-4}
\end{align}
where $\bar{\omega}_a=\sqrt{1+\epsilon_w^2}m_a(t_d)$ is an average of the energy of radiated axions [see Eq.~(\ref{eq4-12})].

The above expression does not depend on $t_d$ except the factor $1/m_a(t_d)$.
However, since the change in the mass of the axion can be negligible ($|\dot{m}_a/m_a^2|\simeq H/m_a <1$) for $t>t_1$,
we can approximate $m_a(t_d)\approx m_a(t_1)$.
Then, the present energy density of axions radiated after $t_1$ is given by
\begin{equation}
\rho_{a,\mathrm{dec}}(t_0) = m_a(0)n_{a,\mathrm{dec}}(t_0) =  \frac{m_a(0)}{\sqrt{1+\epsilon_w^2}m_a(t_1)}\left[{\cal A}_1\frac{\sigma_{\mathrm{wall}}(t_1)}{t_1}+\xi_1\frac{\mu_{\mathrm{string}}(t_1)}{t_1^2}\right]\left(\frac{R(t_1)}{R(t_0)}\right)^3. \label{eq5-3-5}
\end{equation}
The density parameter of axions radiated from the decay of defects is given by
\begin{equation}
\Omega_{a,\mathrm{dec}}h^2 = 8.46\times10^{-2}\times\frac{13.8{\cal A}_1+217\xi_1}{\sqrt{1+\epsilon_w^2}}\left(\frac{g_{*,1}}{70}\right)^{-(n+2)/2(n+4)}\left(\frac{F_a}{10^{12}\mathrm{GeV}}\right)^{(n+6)/(n+4)}\left(\frac{\Lambda}{400\mathrm{MeV}}\right). \label{eq5-3-6}
\end{equation}
As we discussed in Sec.~\ref{sec4}, we use the conservative estimations $\xi_1\simeq 1.0\pm0.5$ and ${\cal A}_1\simeq0.50\pm0.25$. 
Substituting these values and the value of $\epsilon_w$ given by Eq.~(\ref{eq4-11}), we finally obtain
\begin{equation}
\Omega_{a,\mathrm{dec}}h^2 = (5.8\pm2.8)\times\left(\frac{g_{*,1}}{70}\right)^{-(n+2)/2(n+4)}\left(\frac{F_a}{10^{12}\mathrm{GeV}}\right)^{(n+6)/(n+4)}\left(\frac{\Lambda}{400\mathrm{MeV}}\right). \label{eq5-3-7}
\end{equation}
Comparing Eqs.~(\ref{eq5-2-3}) and (\ref{eq5-3-7}), we see that the contribution from domain wall decay is greater than that from string decay.
This result supports the conclusion of Refs.~\cite{1992PhLB..275..279L} and \cite{1994PhRvD..50.4821N}.
\section{\label{sec6}Conclusion}
We have investigated the production of axions from annihilation of domain walls bounded by strings.
We followed the evolution of topological defects by solving classical field equations on the three-dimensional lattice.
In numerical simulations, we observed that global strings indeed annihilate when the mass of the axion becomes relevant.
We calculated the power spectrum of axions produced by the annihilation of string-wall networks by subtracting the contribution which
contains fluctuations given by initial conditions and radiations produced by oscillating strings.
The spectrum has a peak at the low frequency, and the mean energy of radiated axions is $\bar{\omega}_a \simeq 3m_a$,
which is consistent with the result of~\cite{1994PhRvD..50.4821N}.

The total abundance of cold dark matter axions is given by the sum of Eqs.~(\ref{eq5-1-6}), (\ref{eq5-2-3}) and (\ref{eq5-3-7}),
\begin{eqnarray}
\Omega_{a,\mathrm{tot}}h^2 &=& \Omega_{a,0}h^2 + \Omega_{a,\mathrm{str}}h^2 + \Omega_{a,\mathrm{dec}}h^2 \nonumber\\
&=& (8.4\pm3.0)\times\left(\frac{g_{*,1}}{70}\right)^{-0.41}\left(\frac{F_a}{10^{12}\mathrm{GeV}}\right)^{1.19}\left(\frac{\Lambda}{400\mathrm{MeV}}\right), \label{eq6-1}
\end{eqnarray}
where we used $n=6.68$ according to~\cite{2010PhRvD..82l3508W}.
The large uncertainty arises from the poor determination of the scaling parameter $\xi\simeq 1.0\pm0.5$, which might be fixed by
developing the model of the evolution of global string networks, such as the study given by~\cite{2002PhRvD..65d3514M}.
We require that $\Omega_{a,\mathrm{tot}}h^2$ must not exceed the observed value of the abundance of the cold dark matter $\Omega_{\mathrm{CDM}}h^2=0.11$~\cite{2011ApJS..192...18K}.
This gives an upper bound for the axion decay constant
\begin{equation}
F_a\lesssim(2.0\mathchar`-3.8)\times 10^{10}\mathrm{GeV}, \label{eq6-2}
\end{equation}
if we take $g_{*,1}=70$ and $\Lambda=400$MeV.
This bound is more severe than the result of the previous study $F_a<3\times 10^{11}$GeV~\cite{2011PhRvD..83l3531H},
which is obtained by considering only the abundance of axions radiated by strings.
We note that the other group~\cite{2010PhRvD..82l3508W} already reported another bound $F_a\lesssim 3.2^{+4}_{-2}\times 10^{10}$GeV as severe as obtained here,
although they considered only the contribution of axionic strings.
We believe that this severity would come from the larger scaling parameter $\xi\approx 13$ used in their analysis than our numerical prediction $\xi\approx 1$,
which overestimates the relic abundance of axions radiated by strings.
In this sense, we regard that the axion string constraint is milder than that indicated by Eq.~(\ref{eq6-2}).

As we mentioned in Sec.~\ref{sec1}, there is a lower bound $F_a\gtrsim 10^9$GeV which comes from astrophysical observations.
Combining this lower bound with the bound~(\ref{eq6-2}), we conclude that axion models are constrained into the narrow parameter region $F_a\simeq 10^9$-$10^{10}$GeV,
which corresponds to the axion mass $m_a\simeq 10^{-3}$-$10^{-2}$eV.

We note that our numerical result contains some ambiguities as we discussed in Sec.~\ref{sec4}.
Especially, we choose the unrealistic values of parameters such as $\kappa$, $c_T$, and $c_0$, which determines the magnitude of the axion mass
through Eqs.~(\ref{eq2-1-5}) and (\ref{eq2-1-6}), in order to follow the whole relevant processes within one realization of the numerical simulation.
It is not obvious whether the results are sensitive to the tuning of these theoretical parameters.
This should be tested in future high-resolution numerical studies with larger dynamical ranges.
\begin{acknowledgments}
Numerical computation in this work was carried out at the 
Yukawa Institute Computer Facility. 
This work is supported by Grant-in-Aid for
Scientific research from the Ministry of Education, Science, Sports, and
Culture (MEXT), Japan, No.14102004 and No.21111006 (M.~K.~)  and also by
World Premier International Research Center Initiative (WPI Initiative), MEXT, Japan.
K.~S.~and T.~S.~are supported by the Japan Society for the Promotion of Science (JSPS).
T.~H.~was supported by JSPS Grant-in-Aid for Young Scientists (B)
No.23740186 and also by MEXT HPCI STRATEGIC PROGRAM.
\end{acknowledgments}

\end{document}